\documentclass[USenglish,onecolumn,degruyter]{article}

\usepackage[utf8]{inputenc}
\usepackage[big]{dgruyter}
\usepackage{bm}
\usepackage{amsmath}
\usepackage{amssymb}
\usepackage{amsfonts}
\usepackage{listings}
\usepackage{xcolor}
\usepackage{natbib}

  \newcommand{\indep}{\perp\!\!\!\perp}

\newcommand{\E}{\operatorname*{E}}
\begin{document}

  \title{\huge Untangling Sample and Population Level Estimands in Bayesian Causal Computation}

  \runningtitle{Untangling Sample and Population Level Estimands}

  \author*[1]{Arman Oganisian}

  \affil[1]{Department of Biostatistics, Brown University, Rhode Island, USA; E-mail: arman\_oganisian@brown.edu}


  \begin{abstract}
{Model-based Bayesian inference for sample and population-level causal estimands has been growing in popularity. This literature routinely emphasizes clear specification of the target estimand, however blind implementation of standard computational procedures may implicitly target estimands that differ from the one specified at the outset. This sometimes leads to unwitting conflation of sample and population-level inference. In this paper, we elucidate the differences between sample and population-level inference with respect to identification, modeling, computation, and interpretation. For example, common sample-level estimands require cross-world Bayesian modeling, whereas many (but not all) population-level estimands do not. Similarly, the former requires explicit MCMC sampling of counterfactuals from their joint posterior, whereas the latter typically only requires a posterior distribution over parameters and, perhaps, post-hoc Monte Carlo approximations. We explore these issues across four examples, including with Bayesian nonparametric models, in which ostensibly similar Bayesian computational procedures yield posterior draws of fundamentally different estimands, leading to incorrect inferences. We end with a discussion of common mistakes and factors to consider when choosing an estimand.}
\end{abstract}
  \keywords{Bayesian inference, causal inference, missing data, Bayesian nonparametric, Dirichlet process}
   \classification[MSC]{62C10, 62F15, 62G08}

%

\maketitle
\section{Introduction and Existing Work}

Sample-level causal estimands are comparisons of potential outcomes within a subset of units in the observed data sample \citep{rubin1974}. In contrast, population-level causal estimands are comparisons of average (or other summaries) potential outcomes within the population distribution from which we drew this sample. Identification assumptions, Bayesian computation (e.g. associated Markov Chain Monte Carlo (MCMC)), and required modeling differ fundamentally for these two categories but look deceptively similar.

This similarity can lead to a mismatch between the stated estimand and the output of routine computational procedures, which in turn can lead to materially different and possibly erroneous posterior inferences. In our anecdotal experience from attending conference talks, engaging in peer-review service, and reading published and arXiv-ed papers, these errors range from silently imposing cross-world assumptions about the dependence between potential outcomes for a given unit to unwittingly presenting posterior draws of sample-level estimands as draws of population-level estimands (or vice versa) - which can lead to incorrect inferences.

Such issues have been discussed in earlier work by \citet{Ding2018}, who make insightful connections between missing data methods and causal inference stemming from the fact that the treatment assignment indicator can be seen as a kind of ``missing data'' indicator - indicating which potential outcome(s) is missing. They describe Bayesian inference for sample-level effects (Section 3.1) and population-level effects (Section 3.2). They present conjugate parametric models so that conceptual differences can be highlighted without getting into details of computational implementations. Yet, it is exactly in the use of routine/off-the-shelf Monte Carlo (MC) and MCMC procedures where common errors arise - particularly with flexible, nonparametric models.

Similarly, a recent review by \citet{Li2023} also emphasized the differences between sample and population-level Bayesian inference (see in particular their Section 3.a). While \citet{Li2023} provides, a brief illustration of CATE estimation using BART in Section 4.a, the goal of the paper is more expansive and so they do not delve into sample versus population-level computation with such flexible models.

We contribute to this body of work in a few ways. First, while both \citet{Ding2018} and \citet{Li2023} provide a broad overview of many topics in Bayesian causal inference, our work zooms in on complexities of sample and population-level inference with a particular focus on computation. In this sense, this manuscript partially acts as a technical sequel to these papers. Finally, we have a particular focus on computation with nonparametric methods and cross-world identification assumptions. We also provide four examples, worked out in detail, with implementations provided in \texttt{stan}.

We start with a bivariate normal example similar to Example 3 of \citet{Ding2018} and Example 1 of \citet{Li2023}. However unlike these papers, we use it as an opportunity to highlight differences in computation across different covariate distribution models. Specifically, we compare posterior inferences under Bayesian Bootstrap and the empirical distribution (or mixed average treatment effect (MATE) ) against an oracle model and show that different choices can lead to materially different uncertainty estimates.

In the three sub-sections of Section \ref{sc:more_examples}, we discuss three more elaborate examples. First, we illustrate the differences in ITE, CATE, SATE, and PATE estimation using truncated DPM models in order to move the reader past simple parametric models to ones where where the subtleties of MC integration must be considered within a larger MCMC sampling scheme. The second example discusses inference using DPMs for population-level estimands which \textit{do} require cross-world modeling - lest we leave the reader with the incorrect notion that these are uniquely required for sample-level estimands. These two examples deal with only a single non-identifiable parameter governing cross-world dependence and so sensitivity to inferences can be explored completely by presenting inferences across a range of this parameter. We provide a third example in which there are $K>2$ treatment options with a $K$-variate joint Normal model in which cross-world dependence is governed by $K(K-1)/2$ parameters. This is to emphasize the need to be explicit and wary of such assumptions as assessing sensitivity is more challenging.

Finally, we provide a detailed discussion of common errors made in practice that stem from not appreciating the subtle distinctions between sample and population-level inference, with a detailed critique of the Bayesian g-formula of \citet{keil2018} within the context of the issues raised over the course of the preceding examples. We end with a discussion of considerations for choosing between finite and population estimands.

Our overarching message is as follow: when doing Bayesian causal inference, 1) begin with the causal estimand of substantive scientific interest, 2) articulate clearly the unknown quantities on which it depends. This will vary for sample and population-level estimands. 3) Carry out Bayesian inference by following the strict logic of Bayes' rule to find the marginal posterior distribution of those specific unknowns. This procedure will help ensure that inferences are 1) based on the correct posterior, 2) computation is done correctly, and 3) cross-world modeling assumptions are made explicitly and only as necessary. 

Much has been written \citep{DingLiMiratrix2017,Balakrishnan2025} about when, perhaps, inferences made about a particular population-level estimand based on estimators of sample-level analogue are ``conservative'' or not. These are interesting topics but we explicitly avoid them. Rather we focus on correct implementation -  drawing from the posterior distribution of the intended estimand corresponding to a given model and prior. We do not get into when incorrect implementations may be conservative or aggressive.

\section{Notation and Data Structure}

Throughout this manuscript, we work within the Rubin Causal Model (RCM) \citep{rubin1974} and, inferentially, within the Bayesian framework for doing inference on causal effects outlined in \citet{Rubin1978}. This framework defines causal effects as comparisons of potential outcomes, indexed by treatment options $a\in\mathcal{A}$, in a common population. Until Section \ref{sc:more_examples}, we consider cases with a binary treatment $\mathcal{A}=\{0,1\}$ and sometimes refer to patients with $a=0$ colloquially as ``untreated'' and to those with $a=1$ as ``treated'' though both treatment options may be active.

We consider a setting with an observed continuous, real-valued outcome, $Y_i$, for subject $i=1,2,\dots, n$, a continuous real-valued confounder $L_i$, and a treatment indicator $A_i$. We let $Y_i(a)$ denote the potential outcome for subject $i$ under treatment $a\in \{0,1\}$. Throughout we use capital letters to denote random variables and lowercase to denote realized values. We assume the following: stable unit treatment value assumption (SUTVA), $Y = (1-A)Y(0) + AY(1)$; joint conditional exchangeability, $Y(1), Y(0) \indep A \mid L$; and positivity, $P(A = 1 \mid L = l ) > 0$ for all $l $ for which $f(l)>0$. Under SUTVA, each subject's observed outcome is their \textit{factual} potential outcome $Y_i=Y_i(a_i)$. Similarly, their \textit{counterfactual} potential outcome is missing, which we denote as $Y_i^M = Y_i(1-a_i)$, with collection $\bm Y^M=(Y_1^M,Y_2^M,\dots, Y_n^M)$. The observed data is $D^O = \{ y_i, a_i, l_i \}_{i=1}^n$, which contains the factual potential outcome, $y_i=y_i(a_i)$. 

The model-based perspective we adopt (see Chapter 8 of \citet{imbens2015causal} as well as \citet{Rubin1978}) views the $n$ units in $D^C=D^O \cup Y^M=\{y_i(1), y_i(0), a_i, l_i \}_{i=1}^n$ as row-exchangeable realizations from some population distribution, $f(y(1), y(0), a, l)$. Statistical inference for population-level estimands is done by imposing a model on this complete-data distribution and using $D^O$ to making inferences on the parameters governing it. The model-based approach is in contrast to design or randomization-based perspective which focuses on comparisons of potential outcomes for the fixed set of units in $D^O$. From this perspective, the potential outcome pair for each unit is viewed as fixed, and all randomness stems from variation in treatment assignment, $A$, which decides which of the two potential outcomes is revealed to us and which is missing.

Finally, $f(y(1), y(0), a, l)$ may distribute probability mass across a finite set of $N$ units. In this case, the sampling mechanism by which we select $n$ of the total $N$ units into the treatment groups is without replacement and so the treatment allocation probability across subjects are dependent. To avoid such concerns, as is commonly done, we assume $N=\infty$ so that all references to a ``population distribution'' refers to probability distribution over an infinite population \citep{imbens2015causal}. While this is the standard approach in Bayesian causal inference, randomization-based Bayesian inference for finite populations have also been explored \citep{Leavitt2023}.

\section{Sample vs. Population-level Causal Estimands}
With some exceptions, we will primarily focus on the following estimands throughout which are quite common in point-treatment settings in the literature: 
\begin{enumerate}
    \item \textbf{ITE:} The individual treatment effect (ITE) for subject $i \in\{1,2\dots, n\}$,
    $$  \theta_i(Y_i^M) = Y_i(1) - Y_i(0)  $$ 
    we write it as a function of $Y_i^M$ to emphasize that uncertainty about the ITE is due to uncertainty in subject $i$'s counterfactual - if we only knew the counterfactual, then we could compute the difference using the factual we have to get $\theta_i(Y_i^M)$. That is, uncertainty is due to lack of knowledge about the counterfactual. We will suppress this notational dependence and write $\theta_i$ when this emphasis is unwarranted. Additionally, we may sometimes equivalently write $\theta_i(Y_i^M)=(Y_i-Y_i^M)(2A_i-1)$ to facilitate discussion about the $Y^M$ generally without reference to which treatment is indexing $Y^M$.

    \item \textbf{SATE:} The sample average treatment effect (SATE) is simply the sample average of the ITEs:
    $$ \theta(\bm Y^M) = \frac{1}{n} \sum_{i=1}^n \{ Y_i(1) - Y_i(0) \}$$ 
    again, we write $\theta(\bm Y^M)$ to emphasize that uncertainty in the SATE comes from the full set of missing counterfactuals, but may suppress this notation and write $\theta$.
    \item \textbf{CATE:} The conditional average treatment effect (CATE) 
$$\psi(l, \phi_{Y1}, \phi_{Y0} ) = E[Y(1) \mid L=l; \phi_{Y1}] - E[Y(0) \mid L=l; \phi_{Y0}]$$
This is the difference in potential outcomes averaged in the \textit{subpopulation} of the target with $L=l$. It contrasts the average effect had everyone in this subpopulation got treatment 1 versus 0. Here $\phi_{Ya}$ is the parameter vector governing  the conditional outcome distribution under treatment $a$. We write this as a function of $(\phi_{Y1}, \phi_{Y0})$ to emphasize that all uncertainty about the CATE comes from lack of knowledge about these parameters, but will may suppress this and write $\psi(l)$.
\item \textbf{PATE:} The related population-level average treatment effect (PATE) $\Psi = E[Y(1)] - E[Y(0)]$ - the difference in average outcome across the entire target population had everyone taken treatment $A=1$ versus $A=0$. This is obtained by averaging the CATE over the population distribution of the confounder\footnote{Above, $F(l; \phi_L)$ represents the CDF of the covariate distribution and $dF(l; \phi_L)$ is, loosely speaking, the increment in the CDF. This formalism is necessary to discuss both continuous and discrete models in later sections. If $L$ is continuous, $dF(l; \phi_L)=f(l; \phi_L) dl$ where $f$ represents the density function. If $L$ is discrete, $dF(l; \phi_L)=f(l; \phi_L)$ where $f$ represents the probability mass function. Except where otherwise noted, we assume $L$ is continuous.}:
\begin{equation} \label{eq:std}
    \Psi( \phi_{Y1}, \phi_{Y0}, \phi_L)= \int \Big( E[Y(1) \mid L=l; \phi_{Y1}] - E[Y(0) \mid L=l; \phi_{Y0}] \Big)  dF(l; \phi_L)
\end{equation}
 Here, $\phi_L$ is a parameter vector governing the marginal covariate distribution. Again, the notation $ \Psi( \phi_{Y1}, \phi_{Y0}, \phi_L)$ emphasizes that uncertainty in this quantity stems from uncertainty about the three parameter vectors, but we will simply write $ \Psi$ when emphasis is unwarranted. Note that under the causal identification assumptions, each of the two expectations is $E[Y(a) \mid L=l; \phi_{Ya}]=E[Y \mid A=a, L=l; \phi_{Ya}]$ for $a\in\{0,1\}$. Substituting this equivalence into $\Psi$ is known as the standardization formula. It is a special case of the g-formula \citep{robins1986}. 
\end{enumerate}

The first two estimands are not functions of unknown population-level parameters, but rather functions of unknown unit-level counterfactuals for specific units in our sample. Hence, we call these sample-level estimands. The latter two are purely functions of the unknown parameters governing the population distribution of the data, hence we call these population-level estimands. It can also be conceptually helpful to view these estimands as causal contrasts within units nested in successively larger subsets \citep{Kratina2005} - starting from a sample unit (the ITE), to collections of units in the sample (the SATE), to the larger sub-populations from which these units are drawn from (the CATE), and finally to the overall population that contains these sub-populations (the PATE).

\section{The Joint Posterior Distribution} \label{sc:joint_post}

Making Bayesian inferences for any of these estimands follows from the same procedure we use to make Bayesian inferences for all other problems: we ``analyze data by calculating, via Bayes' theorem, the conditional distribution of unknowns given knowns'' \citep{Rubin1978}. That is, we find the marginal posterior over the unknown quantities in our estimand.

What we know is the observed data, $D^O$. The counterfactuals $Y^M$ are unknown and therefore we do not have $D^C$. We also do not know the parameters governing the joint distribution of the data $f(D^C; \omega)$, denoted by $\omega$. Bayesian inference for all causal estimands is done using the joint posterior distribution which, up to a proportionality constant, is given by Bayes' rule as
\begin{equation}\label{eq:joint_post}
f(\bm y^M, \omega\mid D^O) \propto f(D^C; \omega) f(\omega)
\end{equation}
The first factor is the complete-data distribution while the second is the joint prior distribution. As is commonly done in Bayesian analysis, we assume $\omega=(\phi_A, \phi_Y, \phi_L)$ is comprised of distinct component parameter blocks and factorize the joint distribution of the complete data as follows using the chain rule,
\begin{equation} \label{eq:joint}
    \begin{split}
        f(D^C ; \omega) & = \prod_{i=1}^nf(A=a_i\mid y_i(1), y_i(0), l_i; \phi_A) \cdot f(y_i(1), y_i(0) \mid l_i; \phi_Y) \cdot f(l_i; \phi_L)
    \end{split}
\end{equation}
Note the presence of the bivariate distribution of a given subject's potential outcomes, which is not identifiable even under the usual causal assumptions. Suppose, generally, it is parameterized in terms of $\phi_{Y}=(\phi_{Y1}, \phi_{Y0}, \rho)$, where the first two (identifiable) components govern the induced marginal distribution of each potential outcome whereas the (non-identifiable) third component, $\rho$, governs the dependence between the two potential outcomes. In the Rubin Causal Model \citep{rubin1974,Rubin1976, Rubin1978}, the first factor of \eqref{eq:joint} is referred to as ``the design'' while $f(y(1), y(0), l ;\phi_Y, \phi_L)=f(y(1), y(0) \mid l; \phi_Y) f(l; \phi_L)$ is referred to as ``the science.'' 

We typically use an \textit{a priori} independent family of priors which means the joint prior is $f(\omega) = f(\phi_A) f(\phi_{Y}) f(\phi_L)$ with, similarly, $f(\phi_Y)=f(\phi_{Y1}) f(\phi_{Y0})f(\rho)$. This states, for instance, that prior knowledge about the distinct parameters governing the propensity score, would not change our prior beliefs about the outcome model. Parameterization in terms of such distinct and a priori independent parameters has been called ``separability conditions'' in the missing data literature \cite{molenberghs2014} which we adopt for most of the paper. Taken together, the joint posterior distribution of all unknowns is proportional to 
\begin{equation*}
    \begin{split}
        f(\bm y^M, \omega \mid D^O) \propto & f(\omega) \prod_{i| a_i=1} P(A=1\mid y_i(1), y_i(0), l_i; \phi_A) f(y_i(1), y_i(0) \mid l_i; \phi_{Y}) f(l_i\mid \phi_L) \\
         & \ \ \ \times \prod_{i| a_i=0} P(A=0\mid y_i(1), y_i(0), l_i; \phi_A) f(y_i(1), y_i(0) \mid l_i; \phi_{Y}) f(l_i\mid \phi_L) \\
    \end{split}
\end{equation*}
Now, if $A\indep Y(1), Y(0) \mid L$, then the treatment mechanism becomes $ P(A=a_i\mid y_i(1), y_i(0), l_i; \phi_A)= P(A=a_i\mid l_i; \phi_A)$ so that the joint posterior is further simplified to 
\begin{equation} \label{eq:jointpost}
    \begin{split}
        f(\bm y^M, \omega \mid D^O) \propto & f(\omega) \prod_{i| a_i=1} P(A=1\mid l_i; \phi_A) f(y_i(1), y_i(0) \mid l_i; \phi_{Y}) f(l_i\mid \phi_L) \\
         & \ \ \ \times \prod_{i| a_i=0} P(A=0\mid l_i; \phi_A) f(y_i(1), y_i(0) \mid l_i; \phi_{Y}) f(l_i\mid \phi_L) \\
    \end{split}
\end{equation}
From this general result for the joint posterior, paths diverge depending on which causal estimand is of interest. This is because inference for different estimands are done with different marginals of this joint.

\section{The Many Marginals of the Joint Posterior} \label{sc:marginals}

As identified by the g-formula, $\Psi$, is a function of unknown the parameters governing the marginal potential outcome distributions \textit{and} the covariate distribution. So, inference is based on the marginal posterior
$$ f(\phi_{Y1}, \phi_{Y0}, \phi_L \mid D^O) = \int \int \int f(\bm y^M, \omega \mid D^O) \ d\bm y^M \ d \phi_A \ d\rho $$
where we integrate out the counterfactuals, propensity score model parameters, and potential outcome dependence parameter $\rho$. In contrast, $\theta$ is a function only of counterfactuals, $\bm Y^M$, so we integrate out all other unknowns,
$$ f(\bm y^M \mid D^O) = \int f(\bm y^M, \omega \mid D^O) d\omega $$
Similarly, we make inferences for the ITE via the marginal posterior $ f( y_i(1-a_i) \mid D^O)$ and for the CATE via the marginal posterior $f(\phi_{Y1}, \phi_{Y0} \mid D^O)$. Appendix Sections \ref{app:sate} and \ref{app:pate} compute the integrals above for the SATE and PATE, respectively, to illustrate clearly the differences in identification, modeling, and computation underlying both types of estimands.

The key points we stress in this manuscript are the following:
\begin{enumerate}
    \item \textbf{Interpretation}: sample and population-level estimands have different interpretations. The ITE, $\theta_i$, is not the same as the CATE evaluated at $l_i$, $\psi(l_i)$. The former is the treatment effect for subject $i$ in $D^O$. The latter is the average causal effect within the sub-population of patients who have the same covariate profile as subject $i$.
    \item \textbf{Cross-world modeling}: finite-sample estimands such as the ITE and the SATE typically require cross-world assumptions as the marginal posterior of interest is proportional to the \textit{joint} distribution of a given subjects potential outcome \textit{pair}. For example, some invoke a ``no contamination across imputations'' assumption \citep{Rubin2007b}. This asserts conditional independence of the potential outcomes, implying that $f(y_i(0), y_i(1) \mid l_i; \phi_{Y})= f(y_i(0) \mid l_i; \phi_{Y})f(y_i(1) \mid l_i; \phi_{Y})$ in Equation \eqref{eq:jointpost}. In contrast, under the separability conditions with respect to $\rho$, the PATE and the CATE only require assumptions about the marginals of each potential outcome, since the missing counterfactual is integrated out of the joint posterior. Reliance on cross-world assumptions is highly controversial. While assumptions such as ignorability can be forced to hold by designing and controlling the treatment assignment mechanism in a hypothetical trial, there is no trial that can force cross-world assumptions to hold even in principle \citep{Miles2023}.
    \item \textbf{Uncertainty estimation}: Because different estimands are functionals of different unknowns, the posterior uncertainty in each can be dramatically different. For example, posterior uncertainty about the ITE may be much larger than the CATE. Thus, interpreting posterior draws of the ITE as draws of the CATE, for example, may lead to incorrect inferences for the latter.
    \item \textbf{Modeling requirements}: Different estimands require different models. For example, while the CATE does not require a covariate distribution model (its parameters are integrated out), the PATE \textit{does} require it. This means that targeting different estimands will involve different degrees of model specification burden and, as in the previous bullet, different levels of uncertainty.
    \item \textbf{Computation}: Sample-level Bayesian inference for estimands such as $\theta$ and $\theta_i$ requires an ``imputation'' approach where the missing counterfactual for each subject is updated from the posterior. In contrast, Bayesian inference for estimands such as $\psi$ and $\Psi$ only require a posterior for the parameters and do not require missing counterfactuals. While computing the integral in the PATE is often done via MC simulation of potential outcomes, MC simulations are distinct from proper posterior ``imputations''.
\end{enumerate}

In the following section, we provide an implementation example of these differences using a synthetic data set fitting simple parametric models that demonstrates all four points above. Across the three sub-sections of Section \ref{sc:more_examples}, we discuss implementation with three more elaborate examples. The first illustrates differences in ITE, SATE, PATE, and CATE under flexible truncated Dirichlet Process Mixture (DPM) models. We look at a scenario where explicit Monte Carlo simulation is needed compute posterior draws of causal estimands in order to clearly distinguish these from posterior draws. The second example details DPM estimation for population-level estimands that \textit{do} require cross-world assumptions - but ones encoded in a single non-identifiable parameter, $\rho$. In order to highlight the role of cross-world modeling, we look at a third example in which such models involve a large set of parameters. An ``at-a-glance'' summary of some key takeaways from these examples is given in Table \ref{tb:summary}.

\section{Sample vs. Population Effects: A Simple Example in \texttt{Stan}} \label{sc:example}

Given the subtle differences between the ITE, SATE, CATE, and PATE and the errors that could result in implementation, we provide a concrete example\footnote{\url{https://github.com/stablemarkets/untangle}}. This example mirrors Example 3 of \citet{Ding2018} and Example 1 of \citet{Li2023}. However, we 1) discuss the details of MCMC sampling and MC integration and 2) compare inference under different covariate models as this materially impacts inference. We generate synthetic data from the following data generating process (DGP) for $i=1,2,3, \dots, n$ subjects
\begin{equation*}
    \begin{split}
        l_i &\sim  N(\eta,\tau) \\
        a_i \mid l_i & \sim Ber\big( \text{expit}(\gamma_0 + \gamma_1 l_i ) \big) \\
        \begin{bmatrix}
y_i(1) \\
 y_i(0)
\end{bmatrix} \mid l_i & \sim \text{MVN}_2\Big(\begin{bmatrix}
\beta_{01} + \beta_{11} l_i \\
\beta_{00} + \beta_{10} l_i
\end{bmatrix}, \begin{bmatrix}
\sigma_1^2 & \rho \sigma_1 \sigma_0 \\
\rho \sigma_1 \sigma_0 & \sigma_0^2
\end{bmatrix} \Big)
    \end{split}
\end{equation*}
where $\text{MVN}_2( \mu, \Sigma )$ denotes the bivariate normal distribution with the given mean vector and covariance matrix. Together, all the components define a joint distribution for the complete data from Equation \eqref{eq:joint} - comprised of both the science and the design.
\begin{figure}
    \centering
    \includegraphics[width=.85\linewidth]{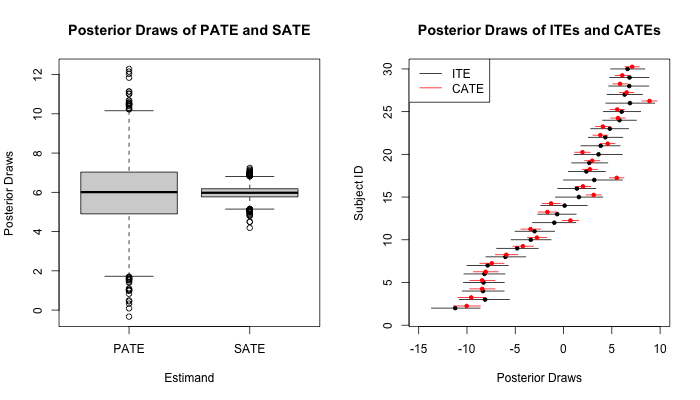}
    \caption{Posterior estimates produced using \texttt{Stan} as described in Appendix Section \ref{app:stan}. Left: boxplot of posterior draws from the distribution of the PATE, $\Psi$, and the SATE, $\theta$. The posterior distributions have the same center, but posterior uncertainty for the PATE is larger. Right: posterior mean (points) and 95\% credible intervals (segments) of each subject's ITE, $\theta_i$, and the CATE evaluated at each $l_i$, $\Psi(l_i)$ for $i=1,2,\dots, 30$. We avoid plotting the last 20 subjects to avoid compression of the plot. The point estimates are similar, but credible intervals in the ITE is wider than the CATE.}
    \label{fig:example}
\end{figure}
To tie in the general notation of the previous sections with the particular model above, $\phi_L=(\eta,\tau)$, $\phi_A = (\gamma_0, \gamma_1)$, $\phi_{Y1}=(\beta_{01}, \beta_{11}, \sigma_1^2)$, $\phi_{Y0}=(\beta_{00}, \beta_{10}, \sigma_0^2)$, and $\phi_Y=(\phi_{Y1}, \phi_{Y0}, \rho)$, where $\rho$ is the correlation coefficient of a given subject's potential outcome. 

The observed outcome is given as $y_i=y_i(1)a_i + y_i(0)(1-a_i)=y_i(a_i)$ and a single generated dataset is given as $D^O = \{ y_i, a_i, l_i\}_{i=1}^n$ with $\bm Y^M$ held out. The true CATE is given by $  \Psi(l) = (\beta_{01} -  \beta_{00}) + ( \beta_{11} - \beta_{10}) l $. Since this model is linear and additive in $L$, it is collapsible and the true PATE is given by $ \Psi = \int \psi(l) f(l; \phi_L) dl = (\beta_{01} -  \beta_{00}) + ( \beta_{11} - \beta_{10}) \eta $ without a need to numerically evaluate the integral.

We simulate data for $n=50$ subjects from the DGP described\footnote{Instead, studies often simulate from and fit outcome models such as 
$$y_i(a) = \beta_0 + \beta_1 a + \beta_2 l_i + \sigma Z_i$$
where $Z_i \sim N(0,1)$ for each $a\in\{0,1\}$. But it is worth noting that this is a very restrictive special case of the DGP used here, even if it has the more familiar regression form. Specifically, it assumes the mean model for potential outcomes $Y(1)$ and $Y(0)$ only differ in their intercept. That is, 1) there are no interaction effects (which our DGP allows for), 2) the conditional variances are the same $\sigma := \sigma_{1}=\sigma_0$, and 3) $\rho=0$. When running formal simulation studies analyzing performance of estimators targeting sample-level estimands, exploring situations where $\rho \neq 0$ would be useful. When running formal simulation studies analyzing performance of estimators targeting population-level estimand, settings with complex interactions and different conditional variances could be insightful.} with covariate parameters $\eta=0$, $\tau=1$, treatment model parameters $\gamma_0=1$ and $\gamma_1=2$, outcome mean parameters $(\beta_{01}=10,\beta_{11}=-4)$, $(\beta_{00}=5,\beta_{10}=5)$, and $\sigma_1=1, \sigma_0=1, \rho=0$. Thus, the true CATE for a given $l$ is $\psi(l) = 5-9l$ and the true PATE is $\Psi = 5$. Under correctly specified models we can make inferences using these data from the following joint posterior (as discussed in Section \ref{sc:marginals}),
\begin{equation*}
    \begin{split}
        f(\bm y^M, \phi_Y, \phi_L \mid D^O) \propto \ & f(\phi_Y, \phi_L) \prod_{i| a_i=1} \text{MVN}_2( y_i,y_i(0) \mid l_i; \phi_{Y}) f(l_i;\phi_L) \\
         & \ \ \ \times \prod_{i| a_i=0} \text{MVN}_2(y_i(1), y_i \mid l_i; \phi_{Y}) f(l_i;\phi_L) \\
    \end{split}
\end{equation*}
where $\text{MVN}_2( \bullet ; -)$ represents the corresponding density function with evaluation vector in $\bullet$ and parameters in $-$. Initially, we fit a correctly specified covariate model so that $f(l; \phi_L) =N(l; \phi_L)$, denoting the normal density function with parameters $\phi_L$ evaluated at $l$. Since we seek a joint posterior over the counterfactuals in addition to the parameters, we must model the joint distribution of the potential outcomes. The parameter $\rho$ in our model governs cross-world dependence. Note however that the CATE is not a function of $\rho$ at all and, therefore, neither is the PATE. Thus, these quantities are not sensitive to prior beliefs about $\rho$. For the example analysis, we use $ f(\phi_Y, \phi_L) \propto U(\rho;-.9,.9) $ - implicitly the prior contributions for all other parameters are $\propto 1$. An alternative would be invoking ``no contamination across imputation,'' which is encoded via a point-mass prior for $\rho$ at zero.

For this particular model we could derive the posteriors exactly, but in more general models we must resort to MCMC methods. Therefore, unlike previous discussions \citep{Li2023,Ding2018}, we use \texttt{Stan} for the MCMC computation. Importantly, since $\bm y^M$ is another unknown along with the parameters, properly generating draws from the joint posterior of all unknowns in \texttt{Stan} requires declaring them in the \texttt{parameters} block of the \texttt{Stan} file rather than simply simulating them in the \texttt{generated quantities} block. More implementation details are give in Appendix Section \ref{app:stan}. Appendix Section \ref{app:sate} provides the details of a Metropolis-in-Gibbs sampler for obtaining the missing counterfactuals manually, without resorting to \texttt{Stan}. As can be seen in that example, counterfactuals can be updated like any other parameter - by proposing a candidate draw for each subject and accepting/rejecting it with the appropriate probability. For now, suppose our MCMC sampler outputs for us draws from this joint posterior and let superscript $(t)$ denote the $t^{th}$ posterior draw. Then, a posterior draw of the CATE is given by  $\Psi^{(t)}(l) = (\beta_{01}^{(t)} -  \beta_{00}^{(t)}) + ( \beta_{11}^{(t)} - \beta_{10}^{(t)}) l $ while a draw of the PATE is given by $ \Psi^{(t)} = (\beta_{01}^{(t)} -  \beta_{00}^{(t)}) + ( \beta_{11}^{(t)} - \beta_{10}^{(t)}) \eta^{(t)} $. A draw of the ITE for a unit assigned to treatment $a_i$ is given by $\theta^{(t)}_i = (y_i - Y^{M,(t)}_i)\cdot (2a_i - 1)$. Note that all variability comes from the unknown counterfactual - the factual is known \textit{a posteriori} since it is part of $D^O$. A posterior draw of the SATE is obtained as $\theta^{(t)}(\bm Y^{M, (t)}) = \frac{1}{n}\sum_{i=1}^n \theta^{(t)}_i$.  

Figure \ref{fig:example} visualizes the posterior draws of the PATE, SATE, CATEs, and ITEs. We can see in the figure that while the posteriors for the SATE and the PATE have a similar center, they have quite different spreads. Again this reinforces the fact discussed in Section \ref{sc:marginals} that these are different quantities with inferences coming from different marginals of the joint posterior. In this setting, if we were to mistakenly do inference for the PATE using posterior draws of the SATE, our credible intervals would be incorrect. Similarly, we can also see that while draws from $f(\theta_i\mid D^O) $ and $f(\psi(l_i) \mid D^O)$ have similar centers, there is greater posterior uncertainty in the ITEs. \\

\subsection{Flexible Covariate Models and Connections to the MATE}

In the previous example, the integral for the PATE had a closed-form expression in terms of the mean parameter of a Gaussian covariate distribution model, $\eta$. In most cases, especially one with many covariates, strong parametric (e.g. Normality) distributional assumptions are not desirable. One way around this is to use the Bayesian bootstrap (BB) model \citep{Rubin_bootstrap}. \cite{Wang2012} first used the BB for this purpose and it has since become mainstream \citep{daniels_book, Oganisian2021a}. The BB models the covariate distribution as discrete with cumulative distribution function $ F(l ; \phi_L) = \sum_{i=1}^n \phi_{Li}\delta_{l_i < l}(l)$. Here, $\delta_{l_i < l}(l)=1$ if the condition $l_i < l$ is true or zero otherwise and the vector $\phi_L=(\phi_{L1}, \phi_{L2}, \dots, \phi_{Ln})$ is a vector in the $n$-simplex. The model is conjugate under an improper Dirichlet prior leading to a proper posterior $\phi_L \mid D^O \sim Dir(1_n)$, where $1_n$ is the length-$n$ unity vector.

Recall from the previous example that the ``exact'' PATE draw - in the sense of being averaged over a posterior draw of the correctly specified covariate model - was $\Psi^{(t)} = \Big( (\beta_{01}^{(t)} -  \beta_{00}^{(t)}) + ( \beta_{11}^{(t)} - \beta_{10}^{(t)}) \Big) \eta^{(t)}$. In the BB approach, using posterior draw $\phi_L^{(t)}=(\phi_{L1}^{(t)}, \phi_{L2}^{(t)}, \dots, \phi_{Ln}^{(t)}) \sim Dir(1_n)$, the $t^{th}$ posterior draw of $\Psi$ is computed as 
\begin{equation*}
    \begin{split}
        \Psi_{BB}^{(t)} = \int \psi^{(t)}(l) dF(l; \phi_L^{(t)})
        = \sum_{i=1}^n \psi^{(t)}(l)  \phi_{Li}^{(t)} 
        = \sum_{i=1}^n \Big((\beta_{01}^{(t)} -  \beta_{00}^{(t)}) + ( \beta_{11}^{(t)} - \beta_{10}^{(t)})l_i \Big) \phi_{Li}^{(t)}
    \end{split}
\end{equation*}
Where the subscript $BB$ is just to remind us this was averaged over the BB covariate model as opposed to another model and does not indicate a change of estimand from the PATE. Early implementations of Bayesian nonparametric causal inference such as by \citet{Hill01012011} instead used the empirical distribution estimate of the covariate distribution $ \hat F(l) = \sum_{i=1}^n \frac{1}{n}\delta_{l_i < l}(l)$, which is a special case of the BB with $\phi_{Li}=1/n$ held fixed. In this approach, the $t^{th}$ posterior draw of the PATE under this covariate distribution estimate is computed as $\Psi^{(t)}_{MATE} = \frac{1}{n}\sum_{i=1}^n \Big((\beta_{01}^{(t)} -  \beta_{00}^{(t)}) + ( \beta_{11}^{(t)} - \beta_{10}^{(t)})l_i \Big)$. \citet{Ding2018} and \citet{Li2023} noted that this procedure does not produce draws of the PATE, but rather a quantity they call a Mixed Average Treatment Effect (MATE), $\Psi_{MATE}=\int \psi(l) d\hat F(l)$. 

\begin{figure}
    \centering
    \includegraphics[width=.9\linewidth]{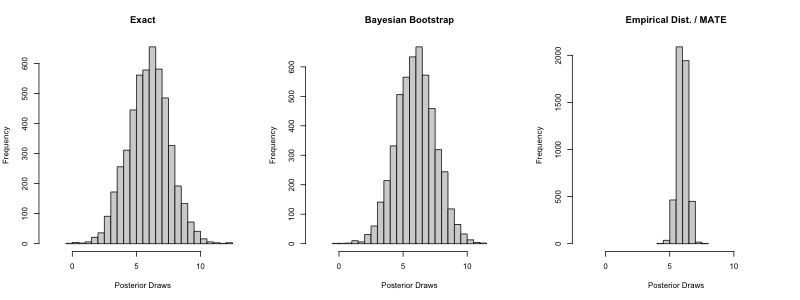}
    \caption{Posterior draws of of the PATE under different covariate models. Left: the ``exact'' draws computed under draws of true covariate model's parameters $\Psi^{(t)} = \Big( (\beta_{01}^{(t)} -  \beta_{00}^{(t)}) + ( \beta_{11}^{(t)} - \beta_{10}^{(t)}) \Big) \eta^{(t)}$. Middle: the BB draws $\Psi_{BB}^{(t)} = \sum_{i=1}^n \psi^{(t)}(l_i) \phi_{Li}^{(t)}$. Right: the MATE draws computed under the empirical distribution held fixed: $\Psi^{(t)}_{MATE} = \frac{1}{n} \sum_{i=1}^n \psi^{(t)}(l_i)$. The BB draws have closer spread to the exact PATE draws - but all share a similar center. }
    \label{fig:example2}
\end{figure}

The posterior draws $\Psi_{MATE}^{(t)}$ will have the same center as draws under the correct covariate model, $\Psi^{(t)}$, and perhaps for this reason it is said that the former can be a ``convenient approximation'' \citep{Li2023} of the latter. However, it's important to note that the spreads of the posterior draws can be quite different due to fixing each $\phi_{Li}= \frac{1}{n}$ and so, in the sense the uncertainty estimation, the approximation may \textit{not} be good. As shown in Figure \ref{fig:example2}, the posterior draws $\Psi^{(t)}_{MATE}$ (right panel) in our example have a much narrower spread than the posterior draws of the $\Psi_{BB}^{(t)}$ (center panel). The spread of the latter is much closer to the posterior of the PATE computed exactly under the correctly specified Normal covariate model, $\Psi^{(t)} = (\beta_{01}^{(t)} -  \beta_{00}^{(t)}) + ( \beta_{11}^{(t)} - \beta_{10}^{(t)}) \eta^{(t)}$ (shown in the left panel).

It is unclear when we would intentionally set out to make inferences about $\Psi_{MATE}$. Rather, as \citet{Ding2018} seem to suggest, since ``we are unwilling to model the possibly multidimensional pretreatment covariate'' we choose to use the empirical distribution and in the process - perhaps inadvertently - end up producing draws of $\Psi_{MATE}$. Our view is that if we truly desire inference for $\Psi_{MATE}$, then we should state that as the target estimand and indeed keep the covariate distribution fixed at the empirical one. However, if we do desire Bayesian inference for the PATE we should proceed as recommended by \citet{daniels_book} (Section 3.2.3) and propagate uncertainty via the BB rather than changing the target from the PATE to the MATE. This follows the generally accepted principle that causal estimands should be chosen based on scientific substance, not modeling or computational convenience. The BB is \textit{a posteriori} centered at the empirical distribution anyway since $E[\phi_{Li} \mid D^O] = \frac{1}{n}$, but also propagates uncertainty. Even frequentist approaches using the empirical distribution typically account for uncertainty in $\hat F$ via the classical Efron bootstrap. The BB is the Bayesian version of this. 

\citet{Li2023} ``view the Bayesian bootstrap as peripheral to causal inference'' because ``how to integrate these samples [of $L$ from $F(l ; \phi_L)$] into the inference of the target causal estimand is case-dependent and generally adds complexity to the analysis compared to the MATE.'' But the MATE is just a special case of the BB that fixes the weights and so the BB is not much more complex. This is because simulated draws of $L$ from $F(l ; \phi_L^{(t)})$ are not required - only draws of the \textit{weights} are needed to compute the weighted sum $\Psi_{BB}^{(t)} = \sum_{i=1}^n \Psi^{(t)}(l_i)  \phi_{Li}^{(t)}$. This is just as with the empirical distribution - a weighted sum with weights $1/n$ is sufficient and we do not need to simulate covariate values with probability $1/n$. This is true for arbitrary CATE model draws $\Psi^{(t)}(l)$ (whether from a generalized linear model or even Bayesian Additive Regression Trees) and Dirichlet draws can be produced in standard software (See Appendix \ref{app:bb}).

Finally, as \citet{Rubin_bootstrap} notes, neither the BB model nor the empirical distribution are a panacea because, while flexible, they are still ``models'' of sorts. They both assume that the true population covariate distribution $f(l)$ has support only on values $l_i$ observed in the sample. In a small sample, if the maximum age we observe is $55$ years, both models assume that patients older than $55$ do not exist in the population. Thus they are essentially misspecified and, if $\psi(l)$ is very different for $l<55$ versus $l>55$, their use may yield incorrect posterior inferences about $\Psi$.

\section{Other Examples}
\label{sc:more_examples}

To solidify ideas and details of correct implementation, in this section we cover three more elaborate examples than the one discussed in Section \ref{sc:example}. First, our discussion up until now did not require discussion of Monte Carlo integration since it was not needed to compute the causal estimands. Thus, Examples 1 and 2 detail situations where numerical integration is required for each posterior parameter draw. Second, our discussion so far may leave one with the mistaken impression that inference for all population-level estimands avoid cross-world assumptions. Example 2 dispels such notions. Finally, dealing with a single cross-world dependence parameter $\rho$ is easy as we can simply present posterior inferences across a range of possible values. Example 3 describes a scenario where the number of such cross-world parameters is growing polynomialy - thus complicating the situation. 

\begin{table}[] 
\centering
\scriptsize
\renewcommand{\arraystretch}{1.25}

\begin{tabular}{lp{4cm}p{4cm}p{4.7cm}}
                                            & Section 6                                                                                 & Section 7.1                                     & Section 7.2                                     \\ \hline
\multicolumn{1}{l|}{Intended Target}        &  ITE, $\theta_i=Y_i(1)-Y_i(0)$                                                            &  PATE, $E[Y(1)-Y(0)]$                                         &  $P(Y(1)>\tau, Y(0)>\tau \mid L=l_i)$                \\ \hline
\multicolumn{1}{l|}{Correct Posterior}      &  $f(y_i^M \mid D^O)$                                                                      &  $f(\phi_Y, \phi_L, \gamma \mid D^O)$                                         &  $f(\phi_Y, \phi_L, \gamma \mid D^O)$                \\ \hline
\multicolumn{1}{l|}{$t^{th}$ Post. Draw}     &  $(\beta_{01}^{(t)} -  \beta_{00}^{(t)}) + ( \beta_{11}^{(t)} - \beta_{10}^{(t)}) l_i $   &  $\frac{1}{n} \sum_{i=1}^n(y_i-Y_i^{M,(t)})(2a_i-1)$ &  $ I(y_i>\tau, Y_i^{M,(t)}>\tau)$   \\ \hline
\multicolumn{1}{l|}{Actual Target}          &  CATE, $\psi(l_i)$                                                                        &  SATE, $\theta(Y^{M})$  &  $ I(Y_i(0)>\tau, Y_i(1)>\tau)$  \\ \hline
\multicolumn{1}{l|}{Problem}                &  The above is a draw of the CATE, $\psi(l_i)$, not the ITE. Interpretation and intervals will differ from those of the ITE.         &  The above is a draw of the SATE, $\theta(Y^{M})$, and is sensitive to cross-world assumptions. Intervals and interpretation differ.                                       &   The above is a draw of a sample level estimand, $ I(Y_i(0)>\tau, Y_i(1)>\tau)$. Intervals and interpretation differ.           \\ \hline
\multicolumn{1}{l|}{Fix}                    &  Transparently impose cross-world assumptions and draw the counterfactual, not the parameters.                                                      &  Draw parameters and conduct MC approximation of integral over $f(l)$. & Draw parameters and conduct MC approximation of integral over $f(y(1), y(0) \mid L=l_i)$.            \\ \hline
\end{tabular}
\caption{Summary of selected key takeaways from examples discussed in Sections 6, 7.1, and 7.2. The first row states the intended target estimand. The second row states the correct marginal posterior that must be drawn from to construct draws of the intended target. The third row shows an incorrectly constructed draw that looks deceptively similar to the target, but is actually a draw of the quantity in the fourth row, ``Actual Target.'' In each case, mistaking the draw in the table as a draw of the intended target may lead to incorrect interpretation and posterior inference. For each case, we summarize the fixes discussed in the text.}
\label{tb:summary}
\end{table}

\subsection{Example 1: CATE vs. ITE with Flexible Bayesian Modeling }

We now consider implementation with a more flexible model that can capture elaborate functional forms in the joint distribution of the potential outcomes. We work with data generated from the following DGP for $i=1,2,3, \dots, n$ subjects
\begin{equation}
    \begin{split}
        l_i &\sim  N(\eta_i, 1) \\
        a_i \mid l_i & \sim Ber\big( \text{expit}(-2 + .1 l_i ) \big) \\
        \begin{bmatrix}
y_i(1) \\
 y_i(0)
\end{bmatrix} \mid l_i & \sim \text{MVN}_2\Big(\begin{bmatrix}
\text{sin}(2l_i) \\
\text{sin}(3l_i)
\end{bmatrix}, \begin{bmatrix}
.1^2 l_i & \rho  (.1\cdot .07) \\
\rho (.1\cdot .07) & .07 ^2 l_i
\end{bmatrix} \Big)
    \end{split}
\end{equation}
where $\eta_1, \eta_2, \dots, \eta_n$ randomly selected from equally-spaced points on the interval $[1,\pi]$. That is, the marginal distribution of $L$ follows a location-mixture distribution. Similarly, the marginal means depend on $l$ through a complicated sine wave rather than the linear and additive functional form used in Example 1. Note that the true CATE in this model is $\Psi(l) = \text{sin}(2l) - \text{sin}(3l)$. We set $\rho=0$ when generating $n=300$ subjects from this process to demonstrate Bayesian inference in the section.

There are many Bayesian models that are commonly seen as ``nonparametric'', ranging from Bayesian Additive Regression Trees (BART) \citep{chipman2010bart} or Dirichlet Processes Mixtures (DPM) \citep{quintana2022dependent} and their truncated analogues \citep{Ishwaran01032001}. See \citet{LineroAntonelli2023} and \citet{Oganisian2021a} for a review - though neither discuss the subtleties of sample- versus population-level inference with such models. The term ``nonparametric'' is a misnomer in Bayesian inference, which always involves parameters. In Bayesian inference, ``nonparametric'' models are better thought of as a spectrum across the ratio of the dimensionality of the parameter space to $n$. With high-dimensional (relative to $n$) models offering more flexibility. Notably, as discussed in detail by \citet{Linero2023}, \citet{Antonelli2022}, and \citet{oganisian_linero2025}, such models require clever ways of relaxing the ``distinct and a priori independent'' parameters assumptions to avoid issues related to ``prior dogmatism'' \citep{Linero2023}. Here, we do not consider such theoretical issues but instead focus strictly on how to back out posterior draws of the sample-level and population-level estimands without conflating them. 

We use a truncated DPM as these can be implemented easily in \texttt{Stan} as opposed to infinite DPMs or BART. Moreover, it lends itself easily to modeling bivariate densities. Both population and sample-level inference can be done with the following DPM model.
\begin{equation*}
    f(y(1), y(0), l; \phi_Y, \phi_L, \gamma) = \sum_{k=1}^K \text{MVN}_2(y(1), y(0) \mid l; \phi_{Y,k}) f(l; \phi_{L,k})  \gamma_k
\end{equation*}
Essentially, the joint distribution of ``the science'' is modeled as a mixture with $K$ components and mixture weights $\gamma=(\gamma_1, \gamma_2, \dots, \gamma_K)$ which live in the $K$-simplex. Each component model is a bivariate normal model $\text{MVN}_2(y(1), y(0) \mid l; \phi_{Y,k})$ governed by parameter collection $\phi_{Y,k}$ containing the parameters of the following mean vector and covariance matrix, respectively:
\begin{equation} \label{eq:mixpars}
\begin{bmatrix}
\beta_{01,k} + \beta_{11,k} l \\
\beta_{00,k} + \beta_{10,k} l
\end{bmatrix} \text{ and } \begin{bmatrix}
\sigma_{1,k}^2 & \rho \sigma_{1,k} \sigma_{0,k} \\
\rho \sigma_{1,k} \sigma_{0,k} & \sigma_{0,k}^2
\end{bmatrix}
\end{equation}
Similarly, we let $f(l; \phi_{L,k})=N(l; \phi_{L,k})$ with $\phi_{L,k}$ containing the mean and variance of the component normal density. We denote by $\phi_Y$ and $\phi_L$ the collection of these parameters across mixture components. Each component model has a mean that is linear and additive in $l$. The idea is that a mixture of simple models for the science can capture more elaborate forms of the science marginally of the mixture components.

In terms of priors, the mixture weights follow a truncated stick-breaking prior process, $f(\gamma)$, \citep{Ishwaran01032001} that induces a ``rich-get-richer'' shrinkage structure. In other words, it generates realizations of $\gamma$ where a few of the $\gamma_k$ sum to $\approx 1$ while the rest are $\approx 0$. That is, even if we conservatively set $K$ to be very large, most of the weight will be on just a few mixture components \textit{a priori}. This is crucial for regularization and reducing sensitivity to choice of $K$. It is similar in spirit to how BART priors shrink towards ``shallow trees.'' The parameters $( \phi_{Y,k}, \phi_{L,k})$, are drawn independently from some specified prior distribution known as a base distribution in this literature. 

Importantly, here, while we allow all other parameters to vary across mixture components, we keep $\rho$ fixed across components. This is because $\rho$ is not identifiable and so trying to learn this parameter in a data-adaptive way will only introduce variability. This ability to allow flexibility for identifiable components of a model while imposing structure on cross-world parameters that are not informed by data is an attractive feature of Bayesian nonparametric inference with DPMs. 

\begin{figure}
    \centering
    \includegraphics[width=1\linewidth]{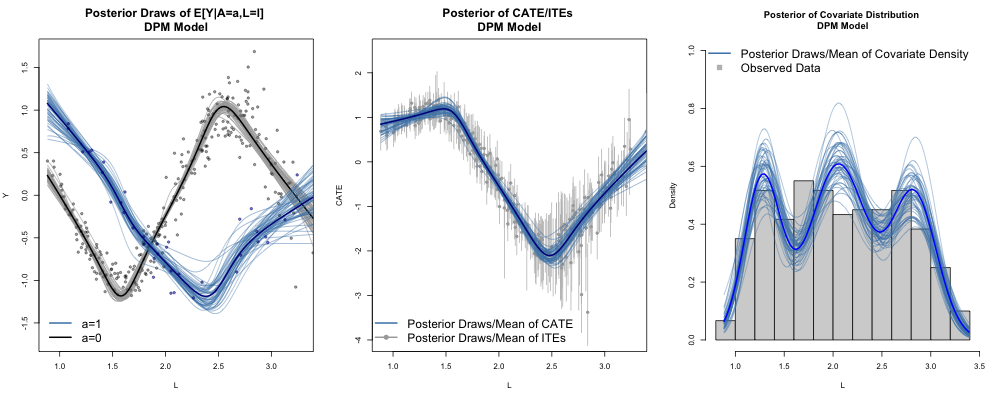}
    \caption{Results from truncated DPM discussed in Example 1. Left: plot of posterior regression function (mean in bold line with some draws in faded lines) of $Y$ on $L$ for each $a\in\{0,1\}$ against training data. Middle: plot of the posterior CATE function $\psi(l)$ with posterior mean in bold and some realizations in faded blue lines. Posterior mean and 95\% intervals of ITEs are shown in gray. Right: posterior mean  density function (bold) along with some density function draws (in faded lines) against observed data (gray bars).}
    \label{fig:dpm_example}
\end{figure}

This model can be specified in \texttt{stan}, but requires careful coding of the mixture components of the likelihood and declaring $Y_i^M$ explicitly as an unknown quantity in the \texttt{parameters} block. If this is done, $\texttt{stan}$ will return a set of draws from the posterior $f(\bm y^M, \omega \mid D^O)$, where $\omega$ refers to the full set of parameters including $ \omega =\{ \phi_{Y,k}, \phi_{L,k}, \gamma_k\}_{k=1}^K$. We denote the $t^{th}$ posterior draw of these parameters with superscripts $(t)$ as before.

In terms of causal inference, some probability calculations show that the mixture model on the science induces the following marginal regression function for each term of the CATE
$$E[Y(a)\mid L=l; \phi_Y, \phi_L,\gamma] = \sum_{k=1}^K w_k(l; \phi_{L}, \gamma) \Big( \beta_{0a,k} + \beta_{1a,k}\cdot l \Big)$$ 
where the mixture weight is $w_k(l; \phi_{L},\gamma) = \frac{\gamma_k f(l; \phi_{L,k})}{\sum_{k'}\gamma_{k'} f(l; \phi_{L,k'})}$. Since $f(l; \phi_{L,k})$ is Gaussian with mean and variance $\phi_{L,k}=(\phi_{L,k,1}, \phi_{L,k,2})$, then $ w_k(l; \phi_{L},\gamma) \propto \gamma_k \exp( -\frac{1}{2\phi_{L,k,2}}(l- \phi_{L,k,1})^2)$. Similarly, the induced marginal covariate density is a mixture of components $N(l; \phi_{L,k})$, $f(l; \phi_L, \gamma) = \sum_{k=1}^K \gamma_k N(l; \phi_{L,k})$. Note that here the parameters are not distinct. Specifically, the parameters governing the marginal covariate density $\phi_L$ also appear in the marginal outcome regression function. A component regression function gets larger weight if $l$ is close to that component's covariate center in squared distance. In this way, the regression function captures non-linear, non-additive effects of $L$ even if the component regression functions are linear and additive. This is demonstrated in the top left panel of Figure \ref{fig:dpm_example} where we ran a truncated DPM with $K=10$ using the synthetic dataset. Across draws, only about four components were getting substantial weight due to the rich-get-richer shrinkage.

Thus, having available the $t^{th}$ draw of the parameters, we can obtain the $t^{th}$ draw of the CATE by plugging the parameter draws into the expression above under our bivariate normal model and taking the difference:

$$ \psi^{(t)}(l; \phi_Y^{(t)},\phi_{L}^{(t)}, \gamma^{(t)}) =  \sum_{k=1}^K w_k(l; \phi_{L}^{(t)}, \gamma^{(t)}) \Big( \beta_{01,k}^{(t)} - \beta_{00,k}^{(t)}  + (\beta_{11,k}^{(t)}-\beta_{10,k}^{(t)})\cdot l \Big)$$

As shown in the middle panel of Figure \ref{fig:dpm_example}, the posterior under the DPM can capture complex forms of the CATE. Note there is no use of the imputed $\bm y^{M,(t)}$ even though we have them. To compute a posterior draw of the PATE, the standardization formula dictates that we must average the CATE over the marginal covariate density model. The $t^{th}$ draw of this density model is $f(l; \phi_L^{(t)}, \gamma^{(t)}) = \sum_{k=1}^K \gamma_k^{(t)} N(l; \phi_{L,k}^{(t)})$. The right panel of Figure \ref{fig:dpm_example} visualizes these draws and demonstrates that posterior can capture complex shapes of the covariate distribution.

To compute the $t^{th}$ draw of the PATE, no imputations are necessary. However, due to the complicated mixture forms of these distributions, we must perform the averaging of the CATE over the covariate distribution via Monte Carlo simulations as follows:

\begin{enumerate}
    \item With probability $\gamma_k^{(t)}$, simulate $L^{(b)} \sim N(l; \phi_{L,k}^{(t)})$. Do this for $b=1, 2, \dots, B$ to obtain $\{L^{(b)}\}_{b=1}^B$. 
    \item Now compute the posterior draw of the PATE by averaging the CATE over these simulated values:
    $$ \Psi^{(t)} = \frac{1}{B}\sum_{b=1}^B\psi^{(t)}(L^{(b)}; \phi_Y^{(t)},\phi_{L}^{(t)}, \gamma^{(t)}) $$
\end{enumerate}

In practice, since we have draws of $Y^M$, it may be tempting to avoid the above and simply take $\frac{1}{n} \Big( \sum_{i:a_i=1} (y_i - y_i^{M,(t)}) + \sum_{i:a_i=0} (y_i^{M,(t)} - y_i) \Big)$ to be the $t^{th}$ draw of the PATE or take the difference between $y_i$ and $Y_i^{M,(t)}$ to be a draw of the CATE, $\psi^{(t)}(l_i)$. Both are incorrect as the former is instead a draw of the SATE while the latter is a draw of the ITE. The middle panel of Figure \ref{fig:dpm_example} depicts how different these quantities are both conceptually and in terms of posterior spread. Mistaking draws of the ITE or SATE as draws of the CATE and PATE, respectively, can have serious inferential consequences. Additionally, as discussed, these sample-level quantities have substantively different interpretations and, unlike their population-level counterparts, will be sensitive to $\rho$.

\subsection{Example 2: Flexible Population-Level Inference Requiring Cross-World Assumptions}

We continue with the same DGP and truncated DPM mixture from the preceding example. Supposing $Y$ from the last example is a biomarker with levels above $\tau$ considered to be a ``failure'' while levels of below $\tau$ are considered to be a ``success'' outcome. This could be, say, blood pressure or cholesterol level and that we desire inference for $\psi_1(\phi_{Y}) = P(Y(1) > \tau, Y(0) >\tau \mid L=l)$. That is, $\psi_1$ is the proportion of patients in the target \textit{sub-population} with covariate profile $l$ that would have failed under both interventions. This quantity is an integral over the joint population potential outcome distribution model
$$ \psi_1(l; \phi_Y, \phi_L, \gamma) = 1 - \int_{0}^\tau \int_{0}^\tau f(y(1), y(0) \mid l;\phi_Y, \phi_L, \gamma) \ dy(1) \ dy(0) $$ 
Unlike the CATE in Example 1, $\psi_1$ is a function of $\rho$. Thus, this is an example of a population-level causal estimand which requires cross-world modeling. Also, in the previous example the CATE involved conditional outcome expectations with a known functional form. So the integral involved in the expectation did not need to be approximated numerically. However, the integrals in $\psi_1$ cannot be (easily) expressed in terms of the population parameters. So, we often approximate such integrals via MC.

Since $\psi_1(l; \phi_Y, \phi_L)$ depends on the joint distribution of the potential outcomes, we must work with the induced conditional (on $l$) \textit{joint} model
$$ f(y(1), y(0) \mid l; \phi_Y, \phi_L, \gamma) = \sum_{k=1}^K w_k(l; \phi_{L}, \gamma) \cdot \text{MVN}_2(y(1), y(0) \mid l; \phi_{Y,k}) $$
Recall that $\phi_{Y,k}$ contains the length 2 mean vector and $2\times 2$ covariance matrix parameters governing the $k^{th}$ multivariate normal mixture component from Equation \ref{eq:mixpars} and the mixture weights are  $w_k(l; \phi_{L},\gamma) = \frac{\gamma_k f(l; \phi_{L,k})}{\sum_{k'}\gamma_{k'} f(l; \phi_{L,k'})}$.

\begin{figure}
    \centering
    \includegraphics[width=.9\linewidth]{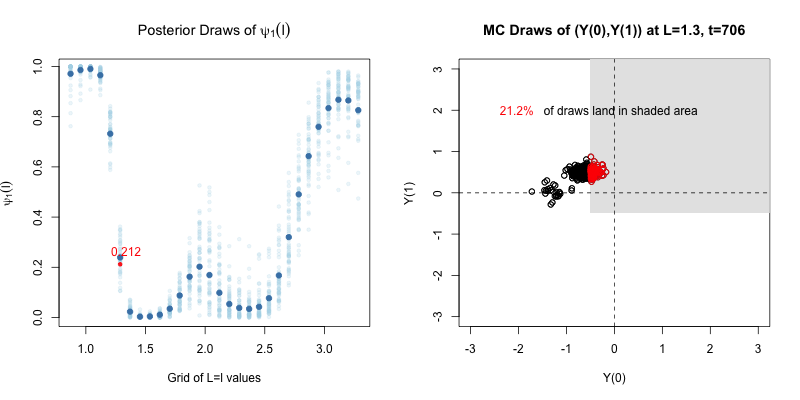}
    \caption{Explanation of MC procedure using synthetic data and truncated DPM model. Left: For $\tau=-.5$, posterior draws of $\psi_1(\phi_{Y}) = P(Y(1) > \tau, Y(0) >\tau \mid L=l)$ at selected grid of $l$ points and posterior mean at each $l$ in bold. Each draw is obtained via MC integration. For instance, the red point at $L=1.3$ is posterior draw at iteration $t=706$. Right: this red point was obtained by simulating $B=500$ draws $[y^{(b)}(0),y^{(b)}(1)]$ from its joint distribution with $t^{th}$ parameter draws plugged in. These are visualized as points. The region of integration $[y(1)>\tau, y(0)>\tau]$ is gray and 19.8\% of the $B$ draws fall in this region.   }
    \label{fig:dpm_example2}
\end{figure}

Again, we can use \texttt{stan} to obtain the $t^{th}$ draw from the posterior of all unknowns $(\bm y^{M,(t)}, \{\phi_{Y,k}^{(t)}, \phi_{L,k}^{(t)},  \gamma_k^{(t)}\}_{k=1}^K  )$. Thus, the $t^{th}$ draw of the bivariate density function is obtained by plugging in these parameters into the expression for $ f(y(1), y(0) \mid l;  \phi_Y, \phi_L, \gamma)$ above and evaluating the double integral at a specified $l$. This can be done as follows in the \texttt{generated quantities} block:
\begin{enumerate}
    \item Compute each weight $ w_k(l; \phi_{L}^{(t)}, \gamma^{(t)})$ for $k=1,2,\dots, K$. These weights necessarily sum to 1.
    \item For $b=1, 2, \dots, B$ simulate $C^{(b)} \in\{1,2,\dots, K\}$ from a multinomial with probability vector given by the weights from the previous step.
    \item For each $b$, simulate a potential outcome pair $\begin{bmatrix}
y^{(b)}(1),
y^{(b)}(0)
\end{bmatrix} \mid L = l \sim \text{MVN}_2(y(1), y(0) \mid l; \phi_{Y, C^{(b)}}^{(t)}) $. These $B$ draws are depicted on the right panel of Figure \ref{fig:dpm_example2} using a synthetic data set at $t=706$ and $l=1.3$.
    \item Compute the $t^{th}$ draw of the causal estimand
    \begin{equation*}
        \begin{split}
            \psi_1^{(t)}(l) & = 1 - \int_{0}^\tau \int_{0}^\tau f(y(1), y(0) \mid l; \phi_Y^{(t)}, \phi_L^{(t)}, \gamma^{(t)}) \ dy(1) \ dy(0) \\
            & \approx \frac{1}{B} \sum_{b=1}^B I( y^{(b)}(1)>\tau, y^{(b)}(0)>\tau )
        \end{split}
    \end{equation*}
    These draws $\psi_1^{(t)}(l)$ across $l$ and $t$ are depicted in the right panel of Figure \ref{fig:dpm_example2} for $\tau=-.5$. The switch from ``$=$'' to ``$\approx$'' reflects that fact that this draw is not calculated via exact evaluation of the integrals, but rather via a Monte Carlo approximation.
\end{enumerate}
That is, we approximate the integrals involved in $\psi_1^{(t)}(l;)$ via $B$ simulated outcome draws from the mixture model of the joint as shown in Figure \ref{fig:dpm_example2}. In general we should take $B$ to be large enough to obtain a desired level of precision at each $t$. This is a controllable computational problem: we can run for successively larger $B$ until results are constant at, say, the third decimal place. The pairs $(y^{(b)}(1), y^{(b)}(0))$ are often incorrectly viewed as ``imputations'' but note that the imputed potential outcomes are $\bm y^{M,(t)}$. Rather the pairs indexed by $b$ are auxiliary simulations generated for MC integration in Step 4. 

The lessons here generalize to many other population-level estimands that require cross-world modeling. For example, the complier-average treatment effect contrasts $E[Y(a) \mid C(1) = C(0)=1]$ across $a\in\{0,1\}$. Here, $C(a)\in\{0,1\}$ is a binary post-treatment indicator of whether someone would have complied with assigned treatment under assignment $a$. This is a population-level quantity - it is the average potential outcome under treatment $a$ among the sub-population that would have complied with treatment under either assignment, $C(1) = C(0)=1$. And yet, it requires cross-world assumptions since it is a integral over the full joint distribution of all four potential outcomes, $\{Y(1), Y(0), C(1), C(0)\}$.

Now, consider an alternative procedure that superficially appears to produce draws of $\psi_1(l; \phi_{Y})$ but does not. Since we have draws $\bm y^{M,(t)}$ from the joint posterior, it may seem natural to then instead combine these with the observed factual to compute $\theta_{1i}^{(t)}(Y^{M,(t)}_i)  = I(y_i>\tau, Y_i^{M,(t)}>\tau)$. Across the $T$ draws, the posterior mean is given by 
$$ P(Y_i(1)>\tau, Y_i(0)>\tau \mid D^O) \approx \frac{1}{T}\sum_{t=1}^T I(y_i>\tau, Y_i^{M,(t)}>\tau)$$ 
Indeed this looks notationally very much like $\psi_1^{(t)}(l) \approx \frac{1}{B} \sum_{i=1}^B I( y^{(b)}(1)>\tau, y^{(b)}(0)>\tau )$. But the crucial distinction is that the probability $P$ in $\psi_1(\phi_{Y})$ is with respect to the population distribution of the bivariate conditional potential outcome distribution. In contrast, the probability $P$ in $P(Y_i(1)>\tau, Y_i(0)>\tau \mid D^O)$ is with respect to the posterior distribution of $Y_i^M$. In fact, $\theta_{1i}^{(t)}(Y^{M,(t)}_i)$ is a draw of the unit-level unknown quantity $\theta_{1i}(Y^{M}_i)  = I(y_i>\tau, Y_i^M>\tau)$ indicating whether subject $i$ would have had a success outcome under either treatment. $P(Y_i(1)>\tau, Y_i(0)>\tau \mid D^O)$ is the posterior expectation of this quantity \textit{for unit $i$}. Thus it has a substantively different interpretation and, because it is a function of missing counterfactuals rather than population parameters, may have very different posterior spreads as we saw with ITEs and CATEs.

\subsection{Example 3: Many Cross-World Parameters }

The requirement of models for cross-world dependence between potential outcomes for some population-level and sample-level estimands may not seem too onerous. After all, we can simply present posterior inference across a range of point-mass priors for $\rho$, or a single prior that averages uncertainty about it as discussed in other work \citep{Ding2018, Li2023}. However, in some settings there may be an unruly amount of cross-world parameters. 

Suppose we are dealing with $K>2$ (note $K$ is redefined from Example 2 now) possible treatment values labeled $\mathcal{A}=\{1,2, \dots, K\}$. In this case, for each subject, we have a potential outcome vector or curve $ \bar Y_i = \{ Y_i(a): a\in\mathcal{A}\}$. We observe only one point on this curve, $y_i(a_i)$, while the other $K-1$ potential outcomes are missing, $Y_i^M = \{ Y_i(a): a\in\mathcal{A}, a\neq a_i\}$. Again the complete data, $D^C = \{\bar Y_i, a_i, l_i\}$ we ideally would like to observe is the union of $D^O=\{y_i(a_i), a_i, l_i\}_{i=1}^n$ and $\{Y_i^M\}_{i=1}^n$. \citet{Balakrishnan2025} tackle the problem of making inferences for the joint distribution of the full potential outcome vector in such settings. Using ideas from optimal transport, they provide methods for conservative inference in the face of non-identifiability of the joint distribution of potential outcomes. We consider a Bayesian solution to this problem - namely Bayesian inference for subject $i$'s curve $\{Y_i(a_i): a_i\in\mathcal{A}\}$. We will show that this is quite different from the population-level analogue $\{ E[Y(a) \mid L=l_i]: a \in\mathcal{A}\}$ and again is sensitive to cross-world parameters.

\begin{figure}
    \centering
    \includegraphics[width=.9\linewidth]{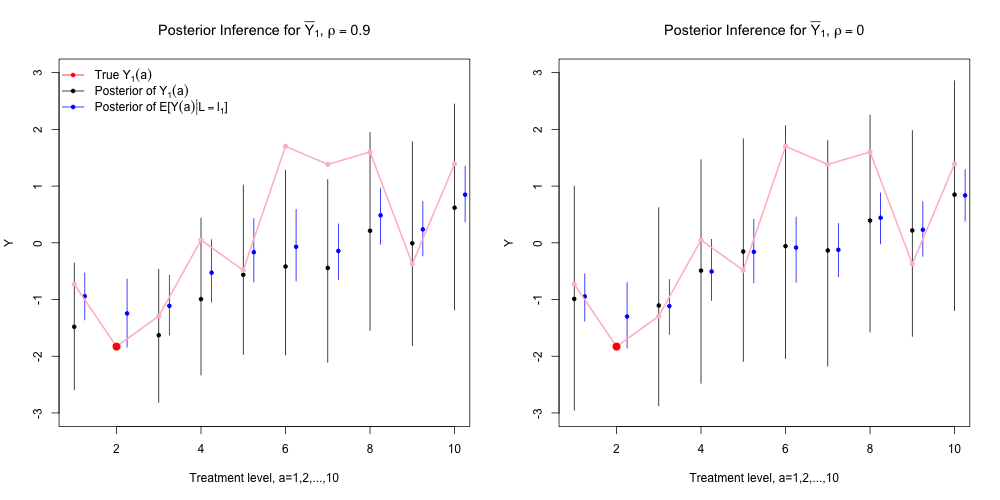}
    \caption{Posterior inference for unit $i=1$'s potential outcome curve across $K=10$ possible treatments using synthetic data under the $K$-variate normal model described in Example 3. Note that $a_1=2$ and so $Y_1(2)$ is observed and plotted as a bold red dot - there is no posterior uncertainty about $Y_1(2)$ since posterior inference is conditional on observed data, which includes $Y_1(2)$. In black are the posterior mean and 95\% credible intervals for $\{Y_1(a_1): a_1\in\mathcal{A}\}$ while the faded red line is the true curve. In blue we show posterior mean and intervals for the population-level analogue $\{ E[Y(a) \mid L=l_i]: a \in\mathcal{A}\}$. The left panel presents inference under $\rho=.9$ and the right panel presents inference under $\rho=0$.}
    \label{fig:po_curve}
\end{figure}

 The set of differences between any two points on the curve can be seen as a multi-treatment version of the ITE. Thus, we would like to sample from the posterior $f(y_i^M \mid D^O)$. From a Bayesian perspective, the posterior derivations and identification assumptions are straight-forward extensions of the binary treatment case where $\mathcal{A}=\{0,1\}$. Carrying this out formally reveals that we require a model for the joint distribution of the $K$ potential outcomes for a given subject, $f(y(1), y(2),\dots, y(K) \mid L=l; \phi_Y)$. One option is the extension of the bivariate Normal model from earlier examples to the $K$-variate normal: $\begin{bmatrix}
Y(1), Y(2), \dots ,Y(K)
\end{bmatrix} \mid L \sim \text{MVN}_K\Big(\begin{bmatrix}
(\beta_{01} + \beta_{11} l),
(\beta_{02} + \beta_{12} l),
\dots ,
(\beta_{0K} + \beta_{1K} l)
\end{bmatrix}, \Sigma \Big) $. Here, $\Sigma$ is the $K\times K$ covariance matrix with element in row $u$ and column $v$ denoted by $\Sigma_{uv}$. While the diagonal elements of $\Sigma$ as well as the marginal mean parameters, $(\beta_{0k}, \beta_{1k})$, are informed by data, the $K(K-1)/2$ off-diagonal elements of $\Sigma$ are not informed by data. In other words, even with this simple model we must make assumptions about a set of cross-world dependence parameters that grows polynomialy in the number of treatments, $K$. Even with $K=4$ this means presenting inference across $6$ correlation parameters across a hypercube $[-1,1]^6$ - clearly infeasible. So we can only proceed under strong, unfalsifiable modeling assumptions. For instance, if $a\in\{1,2,\dots, K\}$ represent dose levels increasing in equal-spaced increments, then one could specify $\Sigma_{uv} = \sigma_u^2 \cdot  \rho^{|u-v|}$. This imposes an autoregressive dependence between potential outcomes along with a homoskedasticity assumption on the marginal variances. It asserts that potential outcomes are more linearly dependent under more similar doses and less dependent under more different doses (in absolute distance). Thus, we reduced to a single non-identifiable parameter $\rho$ but at the cost of a severe autoregressive functional form specification. Under this model Figure \ref{fig:po_curve} shows posterior inference for $\{Y_i(a_i): a_i\in\mathcal{A}\}$ for unit $i=1$. We see that even after the imposition of this structure inference for this curve is sensitive to $\rho$ - with wider 95\% credible intervals under for $\rho=0$ versus $\rho=.9$. 

But there are infinitely many specification. For instance, we could have specified squared distance rather than absolute distance. We could have also specified an exchangeable correlation structure. We cannot reasonably present inference for all of them and, therefore, some may argue it is better to reformulate the estimand if possible. 

Moreover, note in Figure \ref{fig:po_curve} the posterior over the population-level curve $\{ E[Y(a) \mid L=l_i]: a \in\mathcal{A}\}$ has much narrower posterior intervals - again reinforcing the need to not mistake posterior draws of one for the other even if they look like similar quantities.

This issue is also present in time-varying treatment/confounder problems in which a sequence of $K$ treatment decisions are sequentially made over time. Here, the potential outcome is indexed by the whole vector of treatment decisions $Y_i(a_{i1}, a_{i2} \dots, a_{iK})$. If each of the $K$ decisions are binary, then there are exponentially many $2^K$ potential outcomes that may be mutually dependent \citep{richardson2013}. In fact, the amount of cross-world assumptions required for sample-level inference is even larger as even the last covariate value before each treatment $k$ is a potential outcome of all past treatments, $L_{ik}(a_{i1}, a_{i2},\dots, a_{ik-1})$.

This is perhaps why sample-level inference in time-varying treatment problems is extremely rare in the literature. For instance, all methods described in Section 7.b of \citet{Li2023} target population-level estimands. They agree that ``joint modeling approach is rarely used because it quickly becomes intractable'' in $K$. However, we highlight here that the specific intractability in our view lies in the number of non-identifiable cross-world parameters. Models for other aspects of the joint distribution can always be made more flexible using approaches such as the DPM in Example 2.

\section{On the G-Formula Implementation of Keil et al. (2018)}

\cite{keil2018} develop ``a Bayesian version of the potential outcome distribution, the posterior predictive distribution of the potential outcome'', denoted as $f(\tilde y(a) \mid D^O)$ in the notation of our paper. Their goal is to do inference for the PATE, $\Psi$. However, the posterior predictive inference they present is instead for 
$$ \tilde \Psi =E[\tilde Y(1) \mid D^O] - E[\tilde Y(0) \mid D^O] = \int f(\tilde y(1) \mid D^O) d \tilde y(1) - \int f(\tilde y(0) \mid D^O) d \tilde y(0) $$
Their Equation 5 in Appendix I, Section 7.2 presents what they call a ``Bayesian G-Formula'' for each expectation in $\tilde \Psi$
$$  E[\tilde Y(a)\mid D^O] = \int \int \int E[\tilde Y(a) \mid \tilde L=\tilde l; \phi_{Ya}] f_{\tilde L}(\tilde l; \phi_L) f(\phi_{Ya}, \phi_L\mid D^O)\ d \tilde l \ d\phi_{Ya} \ d\phi_L  $$
This approach is frequently cited when doing Bayesian causal inference but it deviates from the approaches described here and elsewhere \citep{Li2023,Ding2018,daniels_book,Oganisian2021a}. First, note that the expectations in $\tilde \Psi$ are over different distributions from those in $\Psi = E[Y(1)] - E[Y(0)]$. The latter are with respect to the marginal \textit{population distributions} of the potential outcomes while the former are with respect to the marginal posterior predictive distribution of each potential outcome. Then $E[\tilde Y(a) \mid D^O]$ has the interpretation of ``The potential outcome we would expect a posteriori \textit{for an arbitrary subject in the target population}''. That is, it is a unit-level quantity, not a population-level quantity like the PATE. While $E[\tilde Y(a) \mid D^O]$ is a unit-level quantity, it is not the ITE because the ITE is the treatment effect of a unit in $D^O$ - not an arbitrary unit. So, $\tilde \Psi$ is not the PATE, ITE, SATE, nor CATE. Since the g-formula, as defined by \citet{robins1986}, is an identification result for population-level quantities, calling this a ``g-formula'' is misleading.

They propose the following computational strategy for evaluating the integrals, which we glean from page 5 of the paper and associated code in the appendix.
\begin{itemize}
    \item First, simulate a covariate value from $ f_{\tilde L}(\tilde l; \phi_L^{(t)})$. Rather than modeling this distribution, they say it is ``often identical to the population by the observed sample, and...can be sampled by taking the empirical distribution of the baseline covariates in the data.'' That is, they randomly choose a covariate vector from the observed set of vectors uniformly with probability $1/n$, with replacement, $n$ times - the same number of times as the sample size. Denote these by $\tilde l^{(t)}_1, \tilde l^{(t)}_2, \dots, \tilde l^{(t)}_n$.
    \item Then for each posterior predictive draw $\tilde l^{(t)}_i$, draw a predictive potential outcome under each treatment $a\in\{0,1\}$ \textit{independently}
    $$ \tilde{y}_i^{(t)}(a) \sim f(\tilde y(a) \mid \tilde L=\tilde l_i^{(t)}; \phi_Y{a}^{(t)}) $$    
    \item Then, they compute the average of the differences across the $n$ simulations:
    $$ \bar E^{(t)}[\tilde Y(1) - \tilde Y(0) \mid D^O] :\approx \frac{1}{n} \sum_{i=1}^n \tilde{y}_i^{(1)}(a) - \tilde{y}_i^{(0)}(0)$$
\end{itemize}
They take this quantity $\bar E^{(t)}[\tilde Y(1) - \tilde Y(0) \mid D^O]$ to be a ``posterior draw'' of the PATE, $\Psi$, what they denote at times as $rd^{(1,0)}$ (rd for risk difference since they deal with binary outcomes). A few points are worth noting about this computational procedure. First, the average of the $n$ simulations $\tilde l^{(t)}$ is, it seems, meant to be a MC approximation of the integral with respect to the posterior predictive density $f_{\tilde L}(\tilde l; \phi_L^{(t)})$. However, because they mistakenly take these for posterior predictive ``imputations'' for each observed subject, they cap the number of MC simulations  at the sample size, $n$. In small sample settings, this may not be sufficient to eliminate MC error. In time-varying treatment settings, with complicated outcome-covariate feedback, it is even more dubious to cap the number of MC iterations to $n$. This may potentially be a factor why, in their Table 2, the Bayesian credible interval is slightly over-covering in small sample settings, but then gets closer to nominal coverage rates in larger sample sizes. Second, this approach does not account for posterior uncertainty in the unknown confounder distribution because it keeps it fixed at the empirical distribution and so it is more akin to a MATE - but not exactly equal to it.

Thus, interpretations and definitions of $\tilde \Psi$ aside, this procedure mixes and matches computational steps for the PATE with computational steps used for the SATE and, in doing so, blurs the distinction between the two. A May 20, 2020 GitHub commit (commit 9235816) seems to indicate the authors subsequently realize the conflation of MC versus posterior predictive simulations. The commit adds the following comments to the code not present in the appendix of their published paper:
\begin{verbatim}
    // note: ...the sample size of the simulated data need not equal N and 
    should often be larger in complex problems;
\end{verbatim}
In a later commit (commit 3d1ed52), they re-code to do \texttt{M=1000} $>n$ simulations per posterior draw. In the same commit, they add the following comment to a code file seeming to acknowledge that their approach has a finite-sample interpretation rather than the population-level interpretation of the PATE identified by Robins' g-formula:
\begin{verbatim}
 // Here, the baseline covariates are fixed across samples, making this
 a 'conditional' or finite-sample causal effect
\end{verbatim}
However, as we demonstrated earlier, the posterior interpretation, uncertainty, and implementation for sample-level estimands differs substantially from the population-level quantity they initially claimed to be inferring. Moreover, if by ``finite-sample causal effect'' they mean they are now inferring the SATE, then this is also incorrect. The procedure neither 1) models the bivariate distribution of subjects' potential outcomes nor 2) declares the missing counterfactuals in the \texttt{parameters} block of their \texttt{stan} file. Instead, they generate them in the \texttt{generated quantities} block. Thus, if their goal was to do inference for the SATE, they are doing so under implicit cross-world ``no contamination across imputations'' assumption \citep{Rubin2007b}.

\section{Discussion} 

Given their subtle yet important differences, should we make inferences on sample or population-level estimands? In certain settings, the estimand of interest is inherently sample-level. For example, \citet{Morales2026} set out to estimate the causal effect of a consumption tax policy on sales in Philadelphia. Since they are interested in estimating the causal effect of this policy in Philadelphia specifically, the estimand in question is sample-level and inference is done via the marginal posterior of Philadelphia's counterfactual sales in absence of the policy. A PATE estimand would be nonsensical as there is no super-population of ``Philadelphias'' from which we drew the one in question. Thus, if the scientific question demands it, then carrying out Bayesian sample-level inference - even with all the associated complications of cross-world modeling - may be preferred over doing inference for a non-existent or ``vaguely defined super-population of study units'' as \citet{Schochet2013} writes in an analysis of the causal effect of four early elementary school math curricula on achievement outcomes. 

Another example of this is the analysis for right-to-carry laws in the United States by \citet{Manski2018}. The unit of observation was at the state level over time with a particular focus on Virginia. They depart from ``conventional practice'' to perform a ``finite-population analysis that views states as the units of interest...to focus attention on the identification problem arising from the unobservability of counterfactual outcomes'' - referencing what we denote with $Y_i^M$. After all, population-level inference for this problem would require positing the existence of some ``random process defined on a superpopulation of alternative nations.'' In a nod to the inherent absurdity, they write that this requires positing ``a random process generating actual American history, with its division of the country into states with their populations of persons, as one among a set of possible histories.''

Similarly, \citet{balzer2016} argue for sample-level estimands like the SATE as it ``remains interpretable if there is no clear super-population from which the study units were selected.'' They write that the SATE is ``readily interpretable as the intervention effect for the sample at hand...'' and ``avoids assumptions about randomly sampling from and generalizing to'' a population.

On the other hand, such sample-level estimands require cross-world assumptions that, as we showed in Section 7.3, may depend on an intractable number of non-identifiable parameters. An assumption like ignorability, $A\indep Y(1), Y(0) \mid L$ is not testable in a given data set. But at least in principle we can benchmark an observational study relying on this assumption to a randomized clinical trial where it holds. If results differ, it may indicate a violation. That is, the assumption is falsifiable. However, cross-world assumptions such as $Y(1) \indep Y(0) \mid L$ are not even falsifiable since we cannot design a trial where independence holds across multiple worlds. Entire approaches based on \textit{single-world} intervention graphics (SWIGs) \citep{richardson2013} exist for identifying causal effects while explicitly avoiding the cross-world assumptions. 

In many applications, a coherent target population of patients \textit{is} well-defined in principle and we are interested in making inferences about this population. When this is the case, we should follow the proper Bayesian procedure for obtaining a posterior distribution over a population-level estimand. As an example, take the analysis of various red blood cell transfusion strategies by \citet{portela_effect_2024}. Here they explicitly state that their target population consists of treatment-eligible patients with myocardial infarction (MI). They state that ``current evidence alone cannot inform transfusion policies in this target population'' and carry out a causal analysis to ``inform the optimal transfusion threshold for this patient population.'' Unlike the right-to-carry example, here it is reasonable to posit the existence of a larger population of patients with MI of which they have only one finite sample of size $n$. Correspondingly, the inverse-weighting procedures they use do target such population-level estimands.

We will not attempt to be prescriptive since the selection of estimands is application specific and depends on the scientific problem of interest. We only advise that one should be explicit about what estimand is being pursued and hope the examples we gave here can help guide correct implementation in either case. When making Bayesian causal inferences, our main recommendation is to - as we have done here - engage in first-principles thinking about the joint posterior distribution of all unknowns and which marginal posterior is actually of interest. Following the strict logic of Bayes' rule and probability calculus will prevent many of the errors of the sort we discussed here. 

\newpage

\vskip 0.2in
\bibliographystyle{plainnat}
\bibliography{sample}

@article{Rubin1976,
   author = {Donald B. Rubin},
   doi = {10.2307/2335739},
   issue = {3},
   journal = {Biometrika},
   month = {12},
   pages = {581},
   title = {Inference and Missing Data},
   volume = {63},
   year = {1976},
}

@article{rubin1974,
  title={Estimating causal effects of treatments in randomized and nonrandomized studies.},
  author={Rubin, Donald B},
  journal={Journal of educational Psychology},
  volume={66},
  doi={10.1037/h0037350},
  number={5},
  pages={688},
  year={1974},
  publisher={American Psychological Association}
}

@article{Leavitt2023,
	title = {Randomization-based, Bayesian inference of causal effects},
	volume = {11},	
	doi = {doi:10.1515/jci-2022-0025},
	number = {1},
	urldate = {2025-12-04},
	journal = {Journal of Causal Inference},
	author = {Leavitt, Thomas},
	year = {2023},
	pages = {20220025},
}

@article{portela_effect_2024,
	title = {Effect of Four Hemoglobin Transfusion Threshold Strategies in Patients With Acute Myocardial Infarction and Anemia},
	volume = {177},
	issn = {0003-4819},
	url = {https://doi.org/10.7326/M24-0571},
	doi = {10.7326/M24-0571},
	number = {11},
	urldate = {2025-12-03},
	journal = {Annals of Internal Medicine},
	author = {Portela, Gerard T. and Carson, Jeffrey L. and Swanson, Sonja A. and Alexander, John H. and Hébert, Paul C. and Goodman, Shaun G. and Steg, Philippe Gabriel and Others},
	year = {2024},
}

@article{Schochet2013,
 ISSN = {10769986, 19351054},
 URL = {http://www.jstor.org/stable/41999423},
 author = {Peter Z. Schochet},
 journal = {Journal of Educational and Behavioral Statistics},
 number = {3},
 pages = {219--238},
 publisher = {American Educational Research Association},
 title = {Estimators for Clustered Education RCTs Using the Neyman Model for Causal Inference},
 urldate = {2025-12-01},
 volume = {38},
 year = {2013}
}

@article{Manski2018,
 ISSN = {00346535, 15309142},
 author = {Charles F. Manski and John V. Pepper},
 journal = {The Review of Economics and Statistics},
 number = {2},
 pages = {pp. 232--244},
 publisher = {The MIT Press},
 title = {How Do Right-to-Carry Laws Affect Crime Rates? Coping with Ambiguity Using Bounded-Variation Assumptions},
 volume = {100},
 year = {2018}
}

@article{DingLiMiratrix2017,
title = {Bridging Finite and Super Population Causal Inference},
author = {Peng Ding and Xinran Li and Luke W. Miratrix},
volume = {5},
number = {2},
journal = {Journal of Causal Inference},
doi = {doi:10.1515/jci-2016-0027},
year = {2017},
lastchecked = {2025-12-03}
}

@article{richardson2013,
  title={Single world intervention graphs (SWIGs): A unification of the counterfactual and graphical approaches to causality},
  author={Richardson, Thomas S and Robins, James M},
  journal={Center for the Statistics and the Social Sciences, University of Washington Series. Working Paper},
  volume={128},
  number={30},
  pages={2013},
  year={2013},
  publisher={Citeseer}
}

@article{balzer2016,
  title={Targeted estimation and inference for the sample average treatment effect in trials with and without pair-matching},
  author={Balzer, Laura B and Petersen, Maya L and van der Laan, Mark J and Search Collaboration},
  journal={Statistics in medicine},
  volume={35},
  number={21},
  doi={10.1002/sim.6965},
  pages={3717--3732},
  year={2016},
  publisher={Wiley Online Library}
}

@article{Rubin1978,
   author = {Donald B. Rubin},
   doi = {10.1214/AOS/1176344064},
   issue = {1},
   journal = {Annals of Statistics},
   month = {1},
   pages = {34-58},
   title = {Bayesian Inference for Causal Effects: The Role of Randomization},
   volume = {6},
   year = {1978},
}

@incollection{Rubin2007b,
    title = {2 Statistical Inference for Causal Effects, With Emphasis on Applications in Epidemiology and Medical Statistics},
    editor = {C.R. Rao and J.P. Miller and D.C. Rao},
    series = {Handbook of Statistics},
    publisher = {Elsevier},
    volume = {27},
    pages = {28-63},
    year = {2007},
    booktitle = {Epidemiology and Medical Statistics},
    doi = {https://doi.org/10.1016/S0169-7161(07)27002-6},
    url ={https://www.sciencedirect.com/science/article/pii/S0169716107270026},
    author = {Donald B. Rubin}
}

@article{Holland1986,
 author = {Paul W. Holland},
 journal = {Journal of the American Statistical Association},
 number = {396},
 pages = {945-960},
 title = {Statistics and Causal Inference},
 urldate = {2024-11-23},
 volume = {81},
 year = {1986}
}

@book{daniels_book,
  title={Bayesian nonparametrics for causal inference and missing data},
  author={Daniels, Michael J and Linero, Antonio and Roy, Jason},
  year={2023},
  publisher={Chapman and Hall/CRC}
}

@article{Rubin_bootstrap,
author = {Donald B. Rubin},
title = {{The Bayesian Bootstrap}},
volume = {9},
journal = {The Annals of Statistics},
number = {1},
publisher = {Institute of Mathematical Statistics},
pages = {130 -- 134},
year = {1981},
doi = {10.1214/aos/1176345338},
URL = {https://doi.org/10.1214/aos/1176345338}
}

@article{linero_agc,
    author = {Linero, Antonio R.},
    title = {Simulation-Based Estimators of Analytically Intractable Causal Effects},
    journal = {Biometrics},
    volume = {78},
    number = {3},
    pages = {1001-1017},
    year = {2021},
    month = {05},
    doi = {10.1111/biom.13499},
    url = {https://doi.org/10.1111/biom.13499},
    eprint = {https://academic.oup.com/biometrics/article-pdf/78/3/1001/54616518/biometrics\_78\_3\_1001.pdf},
}

@article{Wang2012,
    author = {Wang, Chi and Parmigiani, Giovanni and Dominici, Francesca},
    title = {Bayesian Effect Estimation Accounting for Adjustment Uncertainty},
    journal = {Biometrics},
    volume = {68},
    number = {3},
    pages = {661-671},
    year = {2012},
    month = {02},
    doi = {10.1111/j.1541-0420.2011.01731.x}
}

@article{chipman2010bart,
  author = {Hugh A. Chipman and Edward I. George and Robert E. McCulloch},
  title = {{BART}: {B}ayesian additive regression trees},
  volume = {4},
  journal = {The Annals of Applied Statistics},
  number = {1},
  publisher = {Institute of Mathematical Statistics},
  pages = {266 -- 298},
  year = {2010},
  doi = {10.1214/09-AOAS285},
  URL = {https://doi.org/10.1214/09-AOAS285}
}

@article{keil2018,
  title={A Bayesian approach to the g-formula},
  author={Keil, Alexander P and Daza, Eric J and Engel, Stephanie M and Buckley, Jessie P and Edwards, Jessie K},
  journal={Statistical methods in medical research},
  volume={27},
  number={10},
  pages={3183--3204},
  year={2018},
  publisher={SAGE Publications Sage UK: London, England}
}

@article{Oganisian2021a,
   author = {Arman Oganisian and Jason A. Roy},
   doi = {10.1002/SIM.8761},
   issue = {2},
   journal = {Statistics in Medicine},
   month = {1},
   pages = {518-551},
   pmid = {33015870},
   title = {A practical introduction to Bayesian estimation of causal effects: Parametric and nonparametric approaches},
   volume = {40},
   year = {2021},
}

@article{Ding2018,
    author = {Peng Ding and Fan Li},
    title = {{Causal Inference: A Missing Data Perspective}},
    volume = {33},
    journal = {Statistical Science},
    number = {2},
    publisher = {Institute of Mathematical Statistics},
    pages = {214 -- 237},
    year = {2018},
    doi = {10.1214/18-STS645},
}

@article{Li2023,
    author = {Li, Fan  and Ding, Peng  and Mealli, Fabrizia },
    title = {Bayesian causal inference: a critical review},
    journal = {Philosophical Transactions of the Royal Society A: Mathematical, Physical and Engineering Sciences},
    volume = {381},
    number = {2247},
    pages = {20220153},
    year = {2023},
    doi = {10.1098/rsta.2022.0153}
}

@article{Hill01012011,
author = {Jennifer L. Hill},
title = {Bayesian Nonparametric Modeling for Causal Inference},
journal = {Journal of Computational and Graphical Statistics},
volume = {20},
number = {1},
pages = {217--240},
year = {2011},
publisher = {ASA Website},
doi = {10.1198/jcgs.2010.08162}}

@article{Morales2026,
    author = {Fernández-Morales, Esteban and Oganisian, Arman and Lee, Youjin},
    title = {Bayesian shrinkage priors for penalized synthetic control estimators in the presence of spillovers},
    journal = {Biometrics},
    volume = {82},
    number = {2},
    pages = {ujag054},
    year = {2026},
    month = {06},
    doi = {10.1093/biomtc/ujag054}
}

@article{Kratina2005,
author = {Susana Rubin-Bleuer and Ioana Schiopu Kratina},
title = {{On the two-phase framework for joint model and design-based inference}},
volume = {33},
journal = {The Annals of Statistics},
number = {6},
publisher = {Institute of Mathematical Statistics},
pages = {2789 -- 2810},
year = {2005},
doi = {10.1214/009053605000000651},
URL = {https://doi.org/10.1214/009053605000000651}
}

@article{Miles2023,
    author = {Miles, Caleb H},
    title = {On the causal interpretation of randomised interventional indirect effects},
    journal = {Journal of the Royal Statistical Society Series B: Statistical Methodology},
    volume = {85},
    number = {4},
    pages = {1154-1172},
    year = {2023},
    month = {06},
    doi = {10.1093/jrsssb/qkad066}
}

@article{Balakrishnan2025,
title = {Conservative inference for counterfactuals},
author = {Sivaraman Balakrishnan and Edward Kennedy and Larry Wasserman},
pages = {20230071},
volume = {13},
number = {1},
journal = {Journal of Causal Inference},
doi = {10.1515/jci-2023-0071},
year = {2025},
lastchecked = {2025-11-29}
}

@book{imbens2015causal,
  title={Causal inference in statistics, social, and biomedical sciences},
  author={Imbens, Guido W and Rubin, Donald B},
  year={2015},
  publisher={Cambridge university press}
}

@article{Ishwaran01032001,
author = {Hemant Ishwaran and Lancelot F James},
title = {Gibbs Sampling Methods for Stick-Breaking Priors},
journal = {Journal of the American Statistical Association},
volume = {96},
number = {453},
pages = {161--173},
year = {2001},
publisher = {ASA Website},
doi = {10.1198/016214501750332758},
}

@article{oganisian_linero2025,
  title={Priors and Propensity Scores in Bayesian Causal Inference},
  author={Oganisian, Arman and Linero, Antonio},
  journal={Observational studies},
  volume={11},
  number={1},
  pages={47},
  doi={10.1353/obs.2025.a956841},
  year={2025}
}

@article{quintana2022dependent,
  title={The dependent Dirichlet process and related models},
  author={Quintana, Fernando A and M{\"u}ller, Peter and Jara, Alejandro and MacEachern, Steven N},
  journal={Statistical Science},
  volume={37},
  number={1},
  pages={24--41},
  year={2022},
  doi={10.1214/20-STS819},
  publisher={Institute of Mathematical Statistics}
}

@article{Linero2023,
   author = {Antonio R. Linero},
   doi = {10.1080/01621459.2023.2278202},
   journal = {Journal of the American Statistical Association},
   month = {12},
   publisher = {Taylor & Francis},
   title = {In Nonparametric and High-Dimensional Models, Bayesian Ignorability is an Informative Prior},
   year = {2023},
}

@article{Antonelli2022,
   author = {Joseph Antonelli and Georgia Papadogeorgou and Francesca Dominici},
   doi = {10.1111/BIOM.13417},
   issue = {1},
   journal = {Biometrics},
   month = {3},
   pages = {100-114},
   title = {Causal inference in high dimensions: A marriage between Bayesian modeling and good frequentist properties},
   volume = {78},
   year = {2022},
}

@article{LineroAntonelli2023,
    author = {Linero, Antonio R. and Antonelli, Joseph L.},
    title = {The how and why of Bayesian nonparametric causal inference},
    journal = {WIREs Computational Statistics},
    volume = {15},
    number = {1},
    doi = {https://doi.org/10.1002/wics.1583},
    year = {2023}
}

@article{robins1986,
title = {A new approach to causal inference in mortality studies with a sustained exposure period—application to control of the healthy worker survivor effect},
journal = {Mathematical Modelling},
volume = {7},
number = {9},
pages = {1393-1512},
year = {1986},
doi = {https://doi.org/10.1016/0270-0255(86)90088-6},
author = {James Robins},}

@book{molenberghs2014,
  title={Handbook of missing data methodology},
  author={Molenberghs, Geert and Fitzmaurice, Garrett and Kenward, Michael G and Tsiatis, Anastasios and Verbeke, Geert},
  year={2014},
  publisher={CRC Press}
}

\appendix

\section*{\centering Appendix}

\section{Posterior Inference for Sample-Level Estimands}
\label{app:sate}

In this appendix, we take the SATE, $\theta$, as an example and work through the derivation, identification, modeling, and computation to illustrate differences from population-level inference. As discussed, inference for the SATE requires marginalizing out the unknown parameters from the joint posterior to get,

\begin{equation} \label{eq:margpost_SATE}
    \begin{split}
        f(\bm y^M \mid D^O) = & \int f(\bm y^M, \omega \mid D^O) d\omega \\
         \propto  & \int  f(\phi_Y) \prod_{i| a_i=1} f(y_i,y_i(0) \mid l_i; \phi_{Y}) \prod_{i| a_i=0} f(y_i(1), y_i\mid l_i; \phi_{Y}) d\phi_Y \\
    \end{split}
\end{equation}
The second line invokes SUTVA to replace the factual potential outcomes with the observed outcomes - just to make clear which of the two outcomes is a component of the unknowns $\bm Y^M$ and which is observed in $D^O$. The components of $\omega$ governing the treatment and covariate models are all absorbed into a proportionality constant, $C$, as they do not involve $\bm y^M$. Specifically,
$$ C = \int \int \prod_{i=1}^n P(A=a_i \mid l_i, \phi_A) f(l_i;\phi_L) f(\phi_A) f(\phi_L) \ d\phi_A \ d \phi_L $$
Thus, the combination of $A \indep Y(1), Y(0) \mid L$ and imposing prior independence renders the propensity score model ``ignorable'' in the sense of being absorbed into the proportionality constant. See Chapter 5 of \cite{molenberghs2014}) for more on the concept of ``Bayesian ignorability.'' \\

Note that because $\theta$ is a sum of unit-level treatment effects, we must model the full set of parameters governing the entire joint, $\phi_Y=(\phi_{Y1}, \phi_{Y0}, \rho)$, which includes the non-identifiable dependence parameter, $\rho$. Because of the non-identifiability, inference for $\theta$ with a joint model for both potential outcomes is heavily influenced by the prior on $\rho$ \citep{Ding2018,Li2023}. Others discuss an assumption of ``no contamination across imputations'' \citep{Rubin2007b} which asserts conditional independence: $f(y(0), y(1) \mid l)=f(y(0) \mid l)f(y(1) \mid l)$. Under this assumption, it suffices to model the marginals. Note that this assumption that $Y(0) \indep Y(1) \mid L$ is distinct from exchangeability, $A \indep Y(1), Y(0) \mid L$. The latter is an assumption about how the treatment mechanism is related to both potential outcomes and can be satisfied by design. The former is an assumption about how the two potential outcomes are related to each other, and cannot be satisfied be design even in principle. Such modeling assumptions are known as ``cross-world'' assumptions as they impose structure on the relationship between potential outcomes across two hypothetical ``worlds'' in which a subject got treatment $a=1$ and $a=0$. \\

We will typically not be able to compute the integral over $\phi_Y$ in \eqref{eq:margpost_SATE} analytically. Rather, we could proceed by sampling from the posterior using, say, a two-step Metropolis-in-Gibbs Markov Chain Monte Carlo (MCMC) sampling algorithm. At iteration $t$ in the sampler,
\begin{enumerate}
    \item Given previous parameter draws, $\phi_Y^{(t)}$, update each subject's missing potential outcome. For a treated unit, 
    $$ y_i^{(t)}(0) \mid \phi_Y^{(t)}, y_i, l_i \sim \ \propto f( y_i, y_i(0) \mid l_i; \phi_{Y}^{(t)})$$
    and for an untreated unit,
        $$ y_i^{(t)}(1) \mid \phi_Y^{(t)}, y_i, l_i \sim \ \propto f(y_i(1), y_i \mid l_i; \phi_{Y}^{(t)})$$
    These updates may or may not require a Metropolis step depending on the form of the joint potential outcome model. For example, following the usual Metropolis steps, we would propose a candidate value of the counterfactual $y_i^{(t)*}(1)$ from some proposal distribution (which for simplicity we take to be symmetric), then accept or reject this proposal according to how consistent it is with the factual outcome and model parameters via the ratio evaluation 
    $$ \frac{ f(y_i, y_i^{(t)*}(1) \mid l_i; \phi_{Y}^{(t)})}{f(y_i, y_i^{(t-1)}(1) \mid l_i; \phi_{Y}^{(t)})} $$
    After cycling through all units, we have 
    $$\bm Y^{M,(t)} = \big(Y_1^{(t)}(1-a_1), Y_2^{(t)}(1-a_2), \dots, Y_n^{(t)}(1-a_n)\big)$$
    \item Combining the imputation of the missing counterfactuals, $\bm Y^{M,(t)}$, with the observed data $D^O$, update the unknownparameters 

    $$ \phi^{(t)}_Y \mid \bm Y^{M,(t)} \sim \propto  f(\phi_Y) \prod_{i| a_i=1} f( y_i, y_i^{(t)}(0) \mid l_i; \phi_{Y}) \prod_{i| a_i=0} f(y_i^{(t)}(1), y_i \mid l_i; \phi_{Y}) $$
    Again, in general this may require a MH update as the distribution may only be known up to a proportionality constant.
    \item Compute a posterior draw of the SATE

    $$ \theta^{(t)} = \frac{1}{n} \Big( \sum_{i:a_i=1} y_i - y_i^{(t)}(0) + \sum_{i:a_i=0} y_i^{(t)}(1) - y_i \Big)$$
\end{enumerate}
Note that we need not simulate the factual potential outcome - this is observed and fixed \textit{a posteriori} since it is in $D^O$. Only the counterfactual is simulated since it is unknown and not in $D^O$. Across repeated simulations $t=1, 2, \dots, T$, this yields $T$ posterior draws of the SATE which can be summarized via the posterior mean and, say, percentile-based credible intervals. For example, the posterior expectation is approximated via the mean of these draws,
$$ \E[\theta \mid D^O] = \int \theta(\bm y^M) f(\bm y^M \mid D^O) d\bm y^M \approx \frac{1}{T} \sum_{t=1}^T \theta^{(t)}$$
Similarly, since we had to impute each subject's counterfactual in the process, we get inference for the ITE of each subject, $\theta_i$, for free. For example, a point estimate can be formed for the ITE of a treated patient via the posterior mean:
$$ E[\theta_i \mid D^O] = \int \theta_i\big (y_i^M \big) \cdot f(y_i^M\mid D^O) dy_i^M \approx \frac{1}{T} \sum_{t=1}^T y_i - y_i^{(t)}(0)$$
The connections with multiple imputation used in missing-data problems is immediate. The fact that the counterfactual are not in $D^O$ is exactly ``the fundamental problem of causal inference'' \citep{Holland1986}. In this sense, ``we can view causal inference entirely as a missing data problem, where we multiply impute the missing potential outcomes''\citep{Rubin2007b} as in Step 1 of the algorithm above. Specifically, missing data imputation methods such as multiple imputation with chained equations (MICE) iterate between an ``analysis step'' which updates parameters conditional on the complete data and an ``imputation step'' which imputes missing data conditional on those parameters. These steps are analogous to step 2 and step 1, respectively, of the algorithm described.

\section{Posterior Inference for Population-Level Estimands} \label{app:pate}

Now let us contrast the finite-sample inference in the previous appendix section with population-level inference for $\Psi$. Again, we will work through derivation, identification, modeling, and computation to highlight differences. As discussed, inference for this quantity is done by integrating over the unneeded parameters and the missing counterfactuals

\begin{equation*}
    \begin{split}
        f(\phi_{Y1}, \phi_{Y0}, \phi_L \mid D^O) & = \int \int \int f(\bm y^M, \omega \mid D^O) \ d\bm y^M \ d \phi_A \ d\rho \\
        & \propto f(\phi_{Y1})f(\phi_{Y0})f(\phi_{L}) \\
        & \ \ \ \ \times \prod_{i:a_i=1} f(y_i(1) \mid l_i; \phi_{Y1}) f(l_i; \phi_L) \prod_{i:a_i=0} f(y_i(0) \mid l_i; \phi_{Y0})f(l_i; \phi_L)
    \end{split}
\end{equation*}
note that, in the above, the counterfactual for each subject is integrated out of the joint distribution of the potential outcome, leaving just the marginal distribution of the factual outcome of each subject, $f(y_i(a_i) \mid l_i; \phi_{Ya_i})$. This dodges the cross-world assumptions/modeling that was present for the SATE and the ITE.

A constant term $ C = \int \prod_{i=1}^n P(A=a_i \mid l_i; \phi_A) f(\phi_A) d \phi_{A}$
does not involve the parameters of interest, and so was absorbed into the proportionality constant. See \cite{oganisian_linero2025} for a discussion of when the propensity score plays a role in Bayesian causal inference. Since $\Psi$ is also a function of the covariate distribution parameters $\phi_L$, the confounder model \textit{cannot} be absorbed into  the proportionality constant. This is in contrast to posterior inference for the SATE, which relied on a posterior that both required assumptions about the joint distribution of the potential outcomes and did \textit{not} require modeling the covariate distribution.

Models must be specified for the potential outcome under treatment, $f(y(1)\mid l; \phi_{Y1})$, under no treatment $f(y(0)\mid l; \phi_{Y0})$, the confounder model $f(l; \phi_L)$, as well as models for our prior beliefs: $f(\phi_{Y1})$, $f(\phi_{Y0})$, and $f(\phi_{L})$. Typically, the posterior above will not have a known form and instead we must use, as before, Markov Chain Monte Carlo (MCMC) methods to obtain $T$ draws from it, denoted $\{\phi_{Y1}^{(t)}, \phi_{Y0}^{(t)}, \phi_L^{(t)} \}_{t=1}^T$. \\

Given these draws, the $t^{th}$ draw of the PATE is obtained by plugging in each draw into the g-formula
\begin{equation} \label{eq:psi_draw}
    \Psi^{(t)} = \int \Big( E[Y(1) \mid L=l; \phi_{Y1}^{(t)}] - E[Y(0) \mid L=l; \phi_{Y0}^{(t)}] \Big)  f(l; \phi_L^{(t)}) dl
\end{equation}
Now we must compute the integral, a process known as g-computation.\footnote{We often call this ``Bayesian g-computation'', but strictly speaking there are no separate ``frequentist'' and ``Bayesian'' g-formulas. These terms refer to paradigms of statistical inference, while the g-formula is true as a consequence of the tower property and the causal assumptions. It simply maps functionals of the potential outcome distribution to functionals of the observed data distribution. It does not prescribe \textit{how} or within what paradigm we should estimate these observed data functionals.} If the integral is computable analytically, as may be the case with very simple linear/additive regression specifications, we are done. But in complex models deployed in practice, the integral over $L$ cannot be done analytically. Therefore, Monte Carlo (MC) integration is done by simulating, $L^{(1)}, L^{(2)}, \dots, L^{(S)} \sim f(l; \phi_L^{(t)})$ and approximating the integral via an average over these draws as
\begin{equation} \label{eq:psi_approx}
    \Psi^{(t)} \approx \frac{1}{S}\sum_{s=1}^S \Big( E[Y(1) \mid L=L^{(s)}; \phi_{Y1}^{(t)}] - E[Y(0) \mid L=L^{(s)}; \phi_{Y0}^{(t)}] \Big)
\end{equation}
This covers many cases in which standard regression models (e.g. a generalized linear model (GLM)) are used to model each outcome distribution. However, in many other cases, $E[Y(1) \mid L=L^{(s)}; \phi_{Y1}^{(t)}]$ itself may not have a known form into which we can plug in $\phi_{Ya}^{(t)}$. In these cases, we need a second round of MC integration to approximate the conditional expectation at each $L^{(s)}$. That is, we simulate $B$ values $Y^{(1)}(a),Y^{(2)}(a),\dots, Y^{(B)}(a) \sim f(y(a) \mid L=L^{(s)}; \phi_{Ya}^{(t)})$ for each treatment option $a\in\{0,1\}$. Each conditional expectation in \eqref{eq:psi_approx} above is then approximated via an average of these simulations
\begin{equation}
    E[Y(a) \mid L=L^{(s)}; \phi_{Ya}^{(t)}] \approx \frac{1}{B} \sum_{b=1}^B Y^{(b)}(a)
\end{equation}
Note that unlike for the SATE, when targeting the PATE there is no imputation of counterfactuals required since all posterior uncertainty about PATE comes from lack of knowledge about parameters. The simulated $\{ Y^{(1)}(a),Y^{(2)}(a),\dots, Y^{(B)}(a)\}$ and $\{L^{(1)},L^{(1)},, \dots, L^{(S)} \}$ are \textit{not} imputations as in Step 1 of the algorithm in Section \ref{app:sate}. They are just independent draws from the data distribution models done for the purposes of Monte Carlo approximation of expectations. Thus, we should set $B$ and $S$ to be large so that the Monte Carlo approximation error is close to zero. Fast MC integration techniques specially tailored for these causal settings is an active area of research in its own right - see, for example, work by \cite{linero_agc}. For another discussion of this approach to g-computation in a Bayesian framework see Chapter 3.6 \cite{daniels_book} and Section 3.3 of \cite{Oganisian2021a}.\\

Now that we have posterior draws $\Psi^{(t)}$, we often summarize them via point and interval quantities. For example, a posterior mean for $\Psi$ can be approximated as 
$$  E[\Psi \mid D^O] = \int\int\int \Psi( \phi_{Y1}, \phi_{Y0}, \phi_L) f(\phi_{Y1}, \phi_{Y0}, \phi_L \mid D^O)  d\phi_{Y1}d\phi_{Y0} d\phi_{L} \approx \frac{1}{T}\sum_{t=1}^T  \Psi^{(t)} $$
A 95\% credible interval can be formed by taking percentiles of the $T$ draws. \\

Finally, note that posterior inference for the CATE is an automatic by-product of the procedure above - just as the ITE was when inferring the SATE. Specifically, a posterior draw of the CATE at covariate value $l$ is 
\begin{equation} \label{eq:mc_cate}
    \psi^{(t)}(l, \phi_{Y1}^{(t)}, \phi_{Y0}^{(t)} ) = E[Y(1) \mid L=l; \phi_{Y1}^{(t)}] - E[Y(0) \mid L=l; \phi_{Y0}^{(t)}]
\end{equation}
These are the differences we averaged in \eqref{eq:psi_approx} over MC draws of the covariate distribution. The key takeaway here is the Bayesian implementation of the g-formula does not require posterior predictive simulations or additional cross-world assumptions about the joint distribution of the potential outcomes.

\section{Common Implementations and Errors} 
\label{app:errors}

Implementations of Bayesian causal inference sometimes conflate the PATE and the SATE in ways that can lead to incorrect uncertainty estimation and interpretations. This is perhaps because Bayesian causal inference, as popularized initially within the Rubin Causal Model framework, is popularly understood to involve ``imputing the missing counterfactual'' from the posterior predictive. While this is the case for the SATE, this is not the case for the PATE. In this section we walk through some common pitfalls in implementation.

\subsection{Targeting the PATE, but doing ``posterior predictive imputation''}

When attempting to make posterior inferences about the PATE, the following intuitive, yet generally incorrect, computational strategy is sometimes employed. Given posterior draws of the outcome model parameters $\{\phi_{Y1}^{(t)}, \phi_{Y0}^{(t)}\}_{t=1}^T$, a posterior draw of the ``causal effect'' is computed as follows. Given draw $t$,
\begin{enumerate}
    \item For each subject $i$ in the observed data, simulate potential outcome under treatment $a=1$ and $a=0$, 
    $$ y_i^{(t)}(1) \sim f(y(1) \mid l_i; \phi_{Y1}^{(t)}) $$
    $$ y_i^{(t)}(0) \sim f(y(0) \mid l_i; \phi_{Y0}^{(t)}) $$
    these are sometimes referred to as ``posterior predictive draws.''
    \item Average the differences to get 
    $$  \tilde \Psi^{(t)} = \frac{1}{n}\sum_{i=1}^n y_i^{(t)}(1) -  y_i^{(t)}(0)$$
\end{enumerate}
$\tilde \Psi^{(t)}$ is often taken to be a posterior draw of $\Psi$, but this would be incorrect. It departs from the procedure described in Section \ref{app:pate} in two important ways. 

\begin{enumerate}
    \item First, rather than simulating covariates $L^{(1)}, L^{(2)}, \dots, L^{(S)} \sim f(l; \phi_L^{(t)})$, it evaluates the causal effect at each \textit{observed} covariate value. That is, implicitly, it assumes the covariate distribution is the probability mass function that places mass $\phi_{Li}^{(t)}$ on observed covariate value $i$:
    $$ f(l; \phi_L^{(t)}) = \sum_{i=1}^n \phi_{Li}^{(t)} \delta_{l_i}(l)$$
    where the vector of probability masses $(\phi_{Li}^{(t)}, \phi_{L2}^{(t)}, \dots, \phi_{Ln}^{(t)})$ lives in the simplex (i.e. has non-negative elements which sum to 1). The above procedure sets $\phi_{Li}^{(t)}=\frac{1}{n}$, for all $t$ -  thus reducing it to the empirical distribution and ignoring posterior variability.
    \item Second, rather than simulating large $B$ values $Y^{(1)}(a),Y^{(2)}(a),\dots, Y^{(B)}(a) \sim f(y(a) \mid L=L_i; \phi_{Ya}^{(t)})$ for each $L_i$, this approach essentially sets $B=1$ when approximating the CATE at each $l_i$. 
\end{enumerate}

The first departure means that posterior uncertainty (reflected as variation across $t$ in $\tilde \Psi^{(t)}$) does not account for the unknown covariate distribution since it simply simulates from the fixed empirical distribution. This is in contrast to the Bayesian bootstrap approach as discussed in the main text.

The second departure is also concerning as $B=1$ will not in general eliminate Monte Carlo error. Thus variability in $\tilde \Psi^{(t)}$ could be too large as it will capture both posterior variability \text{and} MC error. Credible intervals based on draws of $\tilde \Psi^{(t)}$ may be much too wide, which may lead to over-coverage in repeated samples. However, the average (across $t$) of $\tilde \Psi^{(t)}$ will still be unbiased for the PATE.\\

The error above stems from incorrectly supposing that Bayesian inference for the PATE is a posterior predictive exercise. It is not since the PATE is not a function of the subjects' unknown counterfactuals, but rather a function of unknown parameters.

\subsection{Conflating ITEs and CATEs}

In other settings, analysts target ``the causal effect'' for a given subject - but often fail to specify whether they are referring to the ITE, $\theta_i(Y_i^M)$, or the CATE, $\psi(l, \phi_{Y1}, \phi_{Y0})$, evaluated at $l_i$. As discussed before, these are different quantities with uncertainty stemming from different marginals of the joint posterior, which can lead to different inferences. The ITE is the causal effect for a specific unit in the sample. On the other hand, $\psi(l_i, \phi_{Y1}, \phi_{Y0})$ is average causal effect within the sub-population of units in the target population who have the covariate profile $L=l_i$. \\

As in the previous sub-section, given posterior draws of the outcome model parameters $\{\phi_{Y1}^{(t)}, \phi_{Y0}^{(t)}\}_{t=1}^T$, it may be tempting to compute the posterior draw of the ``causal effect for subject $i$'' as follows: Given draw $t^{th}$ parameter draw,
\begin{enumerate}
    \item Simulate potential outcome under treatment $a=1$ and $a=0$, 
    $$ y_i^{(t)}(1) \sim f(y(1) \mid l_i; \phi_{Y1}^{(t)}) $$
    $$ y_i^{(t)}(0) \sim f(y(0) \mid l_i; \phi_{Y0}^{(t)}) $$
    \item Compute the difference 
    $$  \tilde \theta_i^{(t)} = y_i^{(t)}(1) -  y_i^{(t)}(0)$$
\end{enumerate}

Then, $\tilde \theta_i^{(t)}$ is taken to be a draw of the causal effect ``for subject $i$''. Across draws $t=1,2,\dots, T$, this is believed to yield a set of posterior draws of this effect. If by ``for subject $i$'', the analyst means that $ \tilde \theta_i^{(t)}$ is the $t^{th}$ draw of the ITE, the procedure above is not valid in general. A requirement for validity is the ``no contamination across imputations'' assumption, $Y(1) \indep Y(0) \mid L$, which would justify simulating the two potential outcomes for subject $i$ independently from their marginals. As noted earlier, this is a strong, cross-world assumption, but is seldom mentioned when doing the procedure above. Even under this assumption, however, it is unnecessary to simulate \textit{both} potential outcomes for subject $i$ - one of them is already observed and, therefore, known with certainty. Only $Y_i(1-a_i)$ must be simulated - simulating $Y_i(a_i)$ will lead to extra uncertainty. Thus, this departs from the procedure in Section \ref{app:sate}, which both includes a joint model for the potential outcomes explicitly and imputes only the missing counterfactual.  \\

On the other hand, sometimes the procedure above is done and $\tilde \theta_i^{(t)}$ is mistakenly taken to be the $t^{th}$ draw of the CATE $\psi(l_i, \phi_{Y1}, \phi_{Y0})$. But this falls into the same trap described in the previous section. CATEs are contrasts of \textit{expected} conditional outcomes. We can indeed  view $\tilde \theta_i^{(t)}$ as a Monte Carlo approximation of the conditional expectations in $\psi$, $\tilde \theta_i^{(t)} \approx \psi^{(t)}(l_i, \phi_{Y1}^{(t)}, \phi_{Y0}^{(t)})$ as in equation \eqref{eq:mc_cate}. That is, the simulated draws are taken to be $E[Y(a) \mid L=l_i; \phi_{Ya}^{(t)}] \approx y_i^{(t)}(a)$ with $B=1$ MC iteration for approximation of the expectation. Thus, while the average (across draws $t$) of $\tilde \theta_i^{(t)}$ will be unbiased for $\psi(l_i, \phi_{Y1}, \phi_{Y0})$, the uncertainty will be too large as it will include the Monte Carlo error. If the CATE is truly of interest, then we should take $B>>1$. \\

A common thread between the previous two examples is conflation of ``posterior predictive simulation'' with Monte Carlo approximations of integrals.

\section{\texttt{Stan} code - Joint Posterior Sampling}
\label{app:stan}
Here we provide details of implementing the example discussed in Section 6 in \texttt{Stan}. Specifically, we program a \texttt{Stan} model to sample from the following joint posterior
\begin{equation*}
    \begin{split}
        f(\bm y^M, \phi_Y, \phi_L \mid D^O) \propto \ & f(\phi_Y, \phi_L) \prod_{i| a_i=1} \text{MVN}_2( y_i , y_i(0) \mid l_i; \phi_{Y}) N(l_i\mid \phi_L) \\
         & \ \ \ \times \prod_{i| a_i=0} \text{MVN}_2(  y_i(1), y_i \mid l_i; \phi_{Y}) N(l_i; \phi_L) \\
    \end{split}
\end{equation*}
Importantly, since $\bm y^M$ is another unknown along with the parameters, posterior inference for functionals of these values (such as the ITEs and SATE) requires declaring them in the \texttt{parameters} block of the \texttt{Stan} file as follows:
\begin{verbatim}
... 

parameters {
  // outcome mean vector
  real beta01;
  real beta11;
  real beta00;
  real beta10;
  
  // Full covariance matrix
  real<lower=0> sigma1;
  real<lower=0> sigma0;
  real<lower=-1,upper=1> rho;

  // unobserved potential outcomes
  vector[n] y_m;  
  
  ...

}
...
\end{verbatim}
The ``$\dots$'' represents skipped lines for compactness. In other words, the missing counterfactuals are treated just like parameters. We use the transformed parameters block to compute the bivariate normal mean vector and covariance matrix:

\begin{verbatim}
transformed parameters {
  // outcome mean vector
  vector[n] mu1 = beta01 + beta11 * l;
  vector[n] mu0 = beta00 + beta10 * l;

  // Full covariance matrix
  matrix[2,2] Sigma;
  Sigma[1,1] = square(sigma1);
  Sigma[2,2] = square(sigma0);
  Sigma[1,2] = rho * sigma1 * sigma0;
  Sigma[2,1] = Sigma[1,2];
}
\end{verbatim}

Finally, the complete-data likelihood is specified in the \texttt{model} block of the \texttt{Stan} file:

\begin{verbatim}
model {
  rho ~ uniform(-.9, .9);
  
  // covariate model contribution
  l ~ normal(eta, tau);

  // outcome model contribution
  for (i in 1:n) {
    vector[2] mu_i = [ mu1[i] , mu0[i] ]' ;
    vector[2] y_c; // complete potential outcome - observed with missing

    if (a[i] == 1) {
      y_c = [ y[i] , y_m[i] ]' ;
    } else {
      y_c = [ y_m[i] , y[i] ]' ;    
    }

    y_c ~ multi_normal(mu_i, Sigma);
  }
}
\end{verbatim}

Now that the prior and likelihood have been specified, the \texttt{generated quantities} block of the \texttt{Stan} file can be used to compute draws of the various causal quantities of interest. For example, the ITE and SATE can be computed as 
\begin{verbatim}
generated quantities{
  ...
  real ITEs[n];
  real SATE;
  ...
    
  for( i in 1:n){
    if (a[i] == 1) {
      ITEs[i]  = y[i] - y_m[i];       
    } else {
      ITEs[i]  = y_m[i] - y[i];       
    }
  }
  
  SATE = mean(ITEs);  
}
\end{verbatim}
Similarly, the CATE and the PATE can be computed in this block as well. For the PATE, we compute the integral over $L$ using $S=1000$ MC simulations from the parametric Normal model we specified, as dictated by Equation \eqref{eq:psi_approx}
\begin{verbatim}
generated quantities{
  int S=1000;
  ...
  vector[n] CATE;
  real PATE;
  ... 
  CATE = (beta01 + beta11 * l ) - ( beta00 + beta10 * l );
  
  PATE = 0;
  for (s in 1:S){
    real l_sim = normal_rng(eta, tau);
    PATE += ( (beta01 + beta11 * l_sim ) - ( beta00 + beta10 * l_sim ) );
  }
  PATE = PATE/S; 
}
\end{verbatim}

\section{\texttt{Stan} code - Bayesian Bootstrap versus Empirical Distribution} \label{app:bb}

Here we provide a more detailed version presentation of integrating CATEs over a Bayesian bootstrap draw of the confounder distribution. Recall that Equation \eqref{eq:psi_draw} is
\begin{equation*}
    \Psi^{(t)} = \int \Big( E[Y(1) \mid L=l; \phi_{Y1}^{(t)}] - E[Y(0) \mid L=l; \phi_{Y0}^{(t)}] \Big)  dF(l; \phi_L^{(t)})
\end{equation*}
Note this CDF is a step function with jump size $dF(l; \phi_L)= \phi_{Li}$ at point $l_i$ and jump $dF(l; \phi_L)=0$ at any other value of $l$. It's a very flexible model and considered nonparametric, where ``nonparametric'' means it has a parameter space growing with $n$. The empirical distribution is the special case that plugs in $\phi_{Li} =\frac{1}{n}$,
$$ \hat F(l) = \frac{1}{n}\sum_{i=1}^n I( l \leq l_i) $$ 

A Bayesian, however, would like to account for uncertainty by placing a $Dir(0_n)$ prior on $\phi_L$, where $0_n$ is the $n$-dimensional vector of zeroes. The resulting posterior is $\phi_L \mid D^O \sim Dir(1_n)$, where $1_n$ is the $n$-dimensional vector of ones. Now given a posterior draw $\phi_L^{(t)}$, the integral above simply becomes 
\begin{equation*}
    \Psi^{(t)} = \sum_{i=1}^n\Big( E[Y(1) \mid L=l_i; \phi_{Y1}^{(t)}] - E[Y(0) \mid L=l_i; \phi_{Y0}^{(t)}] \Big)  \phi_{Li}^{(t)}
\end{equation*}
This is because the integral only adds up the jumps in the CDF and, given the discreteness of the model, there are only $n$ jumps - one at each observed covariate value. The integrand is zero elsewhere, thus allowing us to replace the integral with a sum over the $n$ values.

A key point here is that, unlike in the setting described in the main manuscript, there is no Monte Carlo integration required and so there is no need to simulate $\{L^{(1)}, L^{(2)}, \dots, L^{(S)}\}$. When using the Bayesian Bootstrap, we simply evaluate the CATE at each observed $l_i$, then take a weighted sum where the weights are the $\phi_L \mid D^O \sim Dir(1_n)$ draws. When using the empirical distribution, this reduces to an empirical average of the CATEs:
\begin{equation*}
    \Psi^{(t)} = \frac{1}{n} \sum_{i=1}^n\Big( E[Y(1) \mid L=l_i; \phi_{Y1}^{(t)}] - E[Y(0) \mid L=l_i; \phi_{Y0}^{(t)}] \Big)
\end{equation*}
since we set each $ \phi_{Li}^{(t)}=\frac{1}{n}$.\\

All three of these approaches can be done in \texttt{Stan} within the \texttt{generated quantities} block. The Monte Carlo simulation approach was already demonstrated in Section \ref{app:stan}. Below is the code for the Bayesian bootstrap:

\begin{verbatim}
generated quantities{
  ...  
  real PATE_bb;
  ...
  simplex[n] phi_L = dirichlet_rng(rep_vector(1, n));
  ...
  // PATE - Bayesian Bootstrap
  PATE_bb = 0;
  for (i in 1:n){
    PATE_bb += ( (beta01 + beta11 * l[i] ) - ( beta00 + beta10 * l[i] ) )*phi_L[i];
  }
  ...
}
\end{verbatim}
The \texttt{+=} syntax is a sum increment. I.e. \texttt{x += y} is the same as \texttt{x = x + y}. Here is the code for the empirical distribution approach:
\begin{verbatim}
generated quantities{
  ...
  real PATE_ecdf;
  ...
    
  // PATE - Empirical distribution
  PATE_ecdf = 0;
  for (i in 1:n){
    PATE_ecdf += ( (beta01 + beta11 * l[i] ) - ( beta00 + beta10 * l[i] ) )*(1.0/n);
  }

}
\end{verbatim}
Note the decimal format in \texttt{1.0/n} is needed to prevent conversion to integer arithmetic in \texttt{Stan}.

\end{document}